\documentclass[superscriptaddress,onecolumn,showpacs,showkeys,amsmath,aps,prb]{revtex4-1}
\usepackage{graphicx}
\usepackage{dcolumn}
\usepackage[sort&compress]{natbib}
\usepackage{subfigure}
\usepackage{bm}
\usepackage[usenames,dvipsnames,svgnames,table]{xcolor}

\begin{document}

\title{Chebyshev, Legendre, Hermite and other orthonormal polynomials in D-dimensions }
\author{Mauro M. Doria}%
\email{mmd@if.ufrj.br}
\affiliation{Departamento de F\'{\i}sica dos S\'{o}lidos, Universidade Federal do Rio de Janeiro, 21941-972 Rio de Janeiro, Brazil}%
\author{Rodrigo C. V. Coelho}%
\email{rcvcoelho@if.ufrj.br}
\affiliation{ ETH Z\"{u}rich, Computational Physics for Engineering Materials, Institute for Building Materials
Schafmattstrasse 6, HIF, CH-8093 Z\"{u}rich, Switzerland}
\affiliation{Departamento de F\'{\i}sica dos S\'{o}lidos, Universidade Federal do Rio de Janeiro, 21941-972 Rio de Janeiro, Brazil}%

\begin{abstract}
We propose a general method to construct symmetric tensor polynomials  in the D-dimensional Euclidean space which are orthonormal under a general weight. The D-dimensional Hermite polynomials are a particular case of the present  ones for the case of a gaussian weight. Hence we obtain  generalizations of the Legendre and of the Chebyshev polynomials in D dimensions that reduce to the respective well-known orthonormal polynomials in D=1 dimensions. We also obtain new D-dimensional polynomials orthonormal under other weights, such as the Fermi-Dirac, Bose-Einstein, Graphene equilibrium distribution functions and the Yukawa potential. We calculate the series expansion of an arbitrary function in terms of the new polynomials up to the fourth order and define  orthonormal multipoles. The explicit orthonormalization of the polynomials up to the fifth order (N from 0 to 4) reveals an increasing number of orthonormalization equations that matches exactly the number of polynomial coefficients indication the correctness of the present procedure.
\end{abstract}


\date{\today}

\maketitle
\section{Introduction}
The theory of orthonormal polynomials is still an unfolding  branch of Mathematics and Physics~\cite{szego75,gradshteyn14,siafarikas01,chihara01}. It started in the nineteen century and provided the key ingredients for the following century  development of Quantum Mechanics.
Interestingly it was Charles Hermite who firstly introduced tensorial properties to the D-dimensional orthonormal polynomials. For the case of a gaussian weight he obtained symmetric tensorial polynomials and showed them to reduce to the well-known previously found one-dimensional Hermite polynomials by taking the limit $D \rightarrow 1$.
Since then the D-dimensional Hermite polynomials have been studied under several aspects, such as the obtainment of recurrence formulas~\cite{berkowitz70}.
Nevertheless the tensorial orthonormal  polynomials have not been systematically studied for weights other than the gaussian one so far,  apart from a few attempts,  such as the Laguerre weight~\cite{wunsche15} in D=2 with the intent to apply in quantum optics~\cite{wunsche01,kok01}.

In this paper we propose a general method to construct D-dimensional tensorial polynomials orthonormal under an arbitrary weight.
We apply this method  for the first five polynomials (N=0 to 4) and determine their coefficients. The D-dimensional Hermite polynomials are retrieved as the particular case of the gaussian weight. We obtain D-dimensional generalizations of the Legendre and Chebyshev of first and second kind polynomials. By taking the limit $D \rightarrow 1$ the one-dimensional Legendre and Chebyshev of first and second kind polynomials are retrieved.
We choose other weights to construct new D-dimensional tensorial polynomials, such as the Bose-Einstein, the Fermi-Dirac and also the Graphene equilibrium distribution functions. Their interest is in the search for solutions of the Boltzmann equation describing semi-classical fluids~\cite{coelho14, coelho16}. In such cases the corresponding D-dimensional Euclidean space is that of the microscopic velocity. As a last example we construct the D-dimensional polynomials for the Yukawa weight, which are useful in position space to derive the concept of an  orthonormal multipole series expansion.

The D-dimensional Hermite polynomials have many interesting applications in Physics ranging from Quantum Optics~\cite{kok01} to Statistical Mechanics. In the latter case they offer fundamental aid to solve the Boltzmann equation~\cite{grad49b,kremer10, philippi06, coelho16-2} for classical particles. Indeed it was H. Grad who first used the D-dimensional Hermite polynomials  to describe the microscopic velocity space of the Boltzmann equation~\cite{grad49,grad49b, kremer10}. The Boltzmann equation aims  a statistical description of an ensemble of particles and so describes the motion of a set of particles at a scale between the microscopic and the macroscopic levels. While the microscopic level has a deterministic description of motion, since Newton's law is applied to the individual particles, at the macroscopic level the only laws available are those of conservation of mass and momentum for many particles. For fluids and gases the macroscopic level corresponds to the continuity and to the Navier-Stokes equation, respectively, and it can be shown that both follow from the Boltzmann equation~\cite{kremer10, philippi06, coelho14}.
A few decades ago the study of the Boltzmann equation experimented a revival because of a new method developed to solve it on a lattice version of position space. Because of its simplicity this method revolutionized the way to numerically tackle problems in fluid dynamics. It became known as the Lattice Boltzmann Method~\cite{kruger16,succi01} and uses the D-dimensional Hermite polynomials to span the distribution function, which essentially gives the number of particles in a point in phase space. The Gauss-Hermite quadrature is also used in this method to perform integration in the D-dimensional space.

Recently it was found that to render the Lattice Boltzmann applicable to semi-classical fluids the weight that render the D-dimensional polynomials must be the equilibrium distribution function itself. Therefore the gaussian weight is not appropriate because it is associated to the Maxwell-Boltzmann equilibrium distribution function whereas for semi-classical fluids the particles obey the Bose-Einstein or the Fermi-Dirac equilibrium distribution functions\cite{coelho16}. Therefore D-dimensional polynomials orthonormal under these weights are need in such cases and so, one must go beyond the D-dimensional Hermite polynomials.

The remarkably rich tensorial structure of the D-dimensional space is the key element that allows for the existence of the present symmetric tensor polynomials orthonormal under a general weight. This rich tensorial structure was first observed by Harold Grad~\cite{grad49}, but never developed to obtain D-dimensional orthonormal polynomials for a general weight. Here we develop this proposal and obtain the D-dimensional orthonormal polynomials. Notice that this rich tensorial structure is not present for D=1 since the Kronecker's delta function is trivial and equal to one. However in higher dimensions  many tensors can be built as products and sums of the Kronecker's delta function. Harold Grad was the first to notice this wealth of tensors built from Kronecker's delta function~\cite{grad49}.

This paper is organized as follows. In  section~\ref{gen-pol} we propose the general form of the D-dimensional tensorial polynomials and explicitly write the first five ones. In section~\ref{kron} the rich tensorial properties of  D-dimensional space. The explicit construction of the first five (N=0 to 4) orthonormal polynomials is carried in section~\ref{ortho}, which means that all their coefficients are obtained as functions of some integrals over the weight ($I_N$). Next we apply this general theory to specific weights in section~\ref{secweight}.
The known D-dimensional Hermite polynomials are derived from the present ones and also new  D-dimensional generalizations of the Legendre and Chebyshev polynomials of the first and second kinds are proposed here. The projection of such polynomials to D=1 dimensions does give the well-known Hermite in subsection~\ref{dherm}, Legendre (subsection~\ref{dlegen}) and Chebyshev (subsections~\ref{dcheb1} and ~\ref{dcheb2}) which are projected to D=1 dimension (subsection~\ref{proj1d}). Next we consider D-dimensional polynomials orthonormal under new weights, such as Fermi-Dirac (subsection~\ref{fd}), Bose-Einstein (subsection~\ref{be}), graphene (subsection~\ref{graph}) and Yukawa potential (subsection~\ref{yuka}). Finally we  expand a general function in terms of the D-dimensional polynomials in section~\ref{expan}, which leads to the proposal of orthonormal multipoles. We reach conclusions in section~\ref{concl}. Some useful tensorial identities are discussed in appendix~\ref{appendixb}.

\section{General D-dimensional Polynomials}\label{gen-pol}
 Consider the D-dimensional Euclidean space endowed with a weight function $\omega(\boldsymbol{\xi} )$ where the vector  $\boldsymbol{\xi} \equiv (\xi_1,\xi_2,\cdots,\xi_D)$ is defined. We claim here the existence of a set of orthonormal polynomials $\mathcal{P}_{i_1\cdots i_N}( \boldsymbol{\xi})$ in this space.
\begin{flalign}
\int d^D \boldsymbol{\xi} \, \omega( \boldsymbol{\xi} )\mathcal{P}_{i_1\cdots i_N}( \boldsymbol{\xi})\mathcal{P}_{j_1\cdots j_M}(\boldsymbol{\xi})=
\delta_{\scriptscriptstyle {N M}}\delta_{i_1\cdots i_N\vert j_1\cdots j_M}. \label{omeg-feq}
\end{flalign}
The polynomials $\mathcal{P}_{i_1 \cdots i_N}(\boldsymbol{\xi})$ are symmetric tensors in the  D-dimensional Euclidean space, expressed in terms of the vector components $\xi_{i}$ and of $\delta_{ij}$. The  N$^{\mbox{th}}$ order polynomial  is symmetrical in the indices $i_1 \cdots i_N$, and its parity  is $(-1)^N$.
\begin{eqnarray}
 \mathcal{P}_{i_1 \cdots i_N}(-\xi_{i_1},\ldots, -\xi_{i_k},\ldots,-\xi_{i_N})=  (-1)^{N}\mathcal{P}_{i_1 \cdots i_N}(\xi_{i_1},\ldots, \xi_{i_k},\ldots \xi_{i_N})
\end{eqnarray}
The following tensors, defined by Harold Grad~\cite{grad49}, are expressed as sums of products of the Kronecker's delta function, ($\delta_{ij}=1$ for $i=j$ and 0 for $i\neq j$).
\begin{flalign}
\delta_{i_1\cdots i_N\vert j_1\cdots j_N} \equiv \delta_{i_1 j_1}\cdots \delta_{i_Nj_N}\,+\,\mbox{permutations  of  $i$'s},
\label{delta0}
\end{flalign}
and,
\begin{eqnarray}
\delta_{i_1\cdots i_N\, j_1\cdots j_N} \equiv \delta_{i_1 j_1}\cdots \delta_{i_Nj_N}\,+\, \mbox{ all permutations}.
\label{delta1}
\end{eqnarray}
The knowledge of the number of terms in such tensors is useful and  discussed in more details in section~\ref{kron}. The first five (N=0 to 4) polynomials are given by,
\begin{eqnarray}
&&\mathcal{P}_0(\boldsymbol{\xi}) = c_0,\label{H0}
\end{eqnarray}
\begin{eqnarray}
&&\mathcal{P}_{i_1}(\boldsymbol{\xi})=c_1\,\xi_{i_1},\label{H1}
\end{eqnarray}
\begin{eqnarray}
\mathcal{P}_{i_1 i_2}(\boldsymbol{\xi})=c_2\,\xi_{i_1} \xi_{i_2} +f_2(\xi)\,\delta_{i_1 i_2},  \mbox{where} \; f_2(\xi) \equiv {\bar c}_2 \xi^2+ {c^\prime}_2,\label{H2}
\end{eqnarray}
\begin{eqnarray}
\mathcal{P}_{i_1 i_2 i_3}(\boldsymbol{\xi})=c_3 \,\xi_{i_1} \xi_{i_2}\xi_{i_3}  + f_3(\xi)\big (\xi_{i_1} \delta_{i_2 i_3} + \xi_{i_2}\delta_{i_1 i_3} +  \xi_{i_3} \delta_{i_1 i_2}\big),\; \mbox{where} \; f_3(\xi) \equiv {\bar c}_3 \xi^2+ {c^\prime}_3,\label{H3}
\end{eqnarray}
and,
\begin{eqnarray}
&&\mathcal{P}_{i_1 i_2 i_3 i_4}(\boldsymbol{\xi})=c_4 \,\xi_{i_1} \xi_{i_2}\xi_{i_3} \xi_{i_4}+  f_4(\xi)\,\big(\xi_{i_1}\xi_{i_2}
\delta_{i_3 i_4} +\xi_{i_1}\xi_{i_3}
\delta_{i_2 i_4} +\xi_{i_1}\xi_{i_4}
\delta_{i_2 i_3} +\xi_{i_2}\xi_{i_3} \delta_{i_1 i_4}+ \xi_{i_2}\xi_{i_4}
\delta_{i_1 i_3} +\xi_{i_3}\xi_{i_4}\delta_{i_1 i_2} \big ) \nonumber \\
&& + g_4(\xi)\,\delta_{i_1 i_2 i_3 i_4},\; \mbox{where} \; f_4(\xi) \equiv \big( {\bar c}_4 \xi^2+ {c^\prime}_4 \big), \: \mbox{and} \; g_4(\xi) \equiv \big( {\bar d}_4 \xi^4+{d^\prime}_4 \xi^2+  {d}_4 \big).\label{H4}
\end{eqnarray}
Therefore the  N$^{th}$ order polynomial is the sum of all possible symmetric tensors built from products of $\xi_{i}$ and of $\delta_{ij}$ times coefficients which are themselves polynomials in $\xi^2$ to maximum allowed power. This proposal yields a unique expression for the N$^{th}$ order polynomial. \\

 We define integrals $I_{N}$ which are central to the present study. They are assumed to exist and have well defined properties.
\begin{eqnarray}\label{i2n-cont}
I_{N}\,\delta_{i_1\cdots i_{N}} \equiv \int d^D \boldsymbol{\xi} \, \omega( \boldsymbol{\xi} ) \, \xi_{i_1}\cdots \xi_{i_{N}}
\end{eqnarray}
Hereafter the weight function  is assumed to only depend on the modulus of the vector: $\omega(\boldsymbol{\xi} )= \omega(\xi )$, $\xi \equiv\vert \boldsymbol{ \xi} \vert$.
By symmetry it holds that $I_{2N+1}=0$ since the integral vanishes.
Using the spherical integration volume, $\int d^D \boldsymbol{\xi} \, \omega( \boldsymbol{\xi} )= D\pi^{D/2}/\Gamma(D/2+1) \int d\xi\,\xi^{D-1} \omega(\xi) $, the $I_{2N}$ integrals become,
\begin{eqnarray}\label{i2n-cont2}
I_{2N}=\frac{\pi^{\frac{D}{2}}}{2^{N-1}\Gamma\big(N+\frac{D}{2}\big)}\int_{0}^{\xi_{max}}d\xi\, \omega(\xi)\,\xi^{2N+D-1}.
\end{eqnarray}
In case that $\xi_{max} = \infty$  the weight function must have the property $\omega(\boldsymbol{\xi}) \rightarrow 0$ for $ \xi \rightarrow \infty$ faster than any power of $\xi$. Next we shall explicitly prove the orthonormality of the first five polynomials.

\section{D-dimensional tensors based on the Kronecker's delta} \label{kron}
The orthonormality condition of Eq.(\ref{omeg-feq}) shows a rich tensorial structure in D dimensions revealed by the following two important tensors,  $\delta_{i_1\cdots i_N\vert j_1\cdots j_N}$ and $\delta_{i_1\cdots i_N j_1\cdots j_N}$, defined in Eqs.(\ref{delta0}) and (\ref{delta1}), respectively. The former is associated with the orthonormality condition while the latter is the totally symmetric tensor introduced in the definition of the functions $I_N$ given by Eq.(\ref{i2n-cont}). Both tensors are expressed as sums over several terms each one expressed as a product of Kronecker's delta functions.
We determine the number of terms in these two tensors.
The tensor $\delta_{i_1\cdots i_N \vert j_1\cdots j_N}$
has N! terms since this tensor is a sum over all possible permutations of the $i$'s under a fixed set of $j$'s.
The tensor $\delta_{i_1\cdots i_N j_1\cdots j_N}$ has (2N-1)!/2$^{N-1}$(N-1)!=(2N-1)(2N-3)(2N-5)...1 terms according to the arguments below.
Firstly notice that the tensor  $\delta_{i_1\cdots i_N j_1\cdots j_N}$ has more terms than
$\delta_{i_1\cdots i_N \vert j_1\cdots j_N}$ and here we seek to find these remaining  tensors.
\begin{eqnarray}
\delta_{i_1\cdots i_N j_1\cdots j_N} = \delta_{i_1\cdots i_N \vert j_1\cdots j_N} + \mbox{ other tensors}
\end{eqnarray}
Next we determine the remaining ``other tensors'' in case N=0 to 4 and determine the number of components of the above tensor by induction. For this we introduce a short notation that only distinguishes  indices $i$ from $j$ indices.
In this notation the previous expression becomes equal to,
\begin{eqnarray}
\delta_{i i\cdots i \cdots j j \cdots  j}= \delta_{i i\cdots i \vert j j \cdots  j}+ \mbox{ other tensors}.
\end{eqnarray}
For N=1 we have that,
\begin{eqnarray}
\delta_{i j}= \delta_{i\vert j},
\end{eqnarray}
and we express this identity with respect to the number of terms simply as $1 = 1$. For N=2 notice that, $\delta_{i_1 i_2 j_1 j_2}= \delta_{i_1 j_1}\delta_{i_2 j_2}+\delta_{i_1 j_2}\delta_{i_2 j_1}+ \delta_{i_1 i_2}\delta_{j_1 j_2}$.
Since $\delta_{i_1 i_2 \vert j_1 j_2}= \delta_{i_1 j_1}\delta_{i_2 j_2}+\delta_{i_1 j_2}\delta_{i_2 j_1}$ the above relation becomes,
$\delta_{i_1 i_2 j_1 j_2}= \delta_{i_1 i_2 \vert j_1 j_2} + \delta_{i_1 i_2}\delta_{j_1 j_2}$. Therefore it holds that $\delta_{i_1 i_2 j_1 j_2}= \delta_{i_1 i_2 \vert j_1 j_2} + \delta_{i_1 i_2}\delta_{j_1 j_2}$.
While $\delta_{i_1 i_2 j_1 j_2}$ contains 3 components, $\delta_{i_1 i_2 \vert j_1 j_2}$ has only 2, such that it holds for this decomposition that $3 = 2 \oplus 1$. In the short notation the above relation becomes,
\begin{eqnarray}
&& \delta_{i i \vert j j}= \delta_{i j}\delta_{i j} \\
&& \delta_{i i j j}= \delta_{i j}\delta_{i j} + \delta_{i i}\delta_{j j}.
\end{eqnarray}
The tensors $\delta_{i j}\delta_{i j}$ and $\delta_{i i}\delta_{j j}$ contain 2 and 1 components, respectively. Thus the short notation gives that $\delta_{i i\cdots i \vert j j \cdots  j}\equiv \delta_{i j} \delta_{i j} \cdots \delta_{i j}$ where the products of $\delta_{i j}$ takes into account all possible permutations. The tensor $\delta_{i i\cdots i  j j \cdots  j}$ cannot be expressed similarly because not all combinations of $\delta_{i j}$'s are included. Therefore the need to decompose it into $\delta_{i i\cdots i \vert j j \cdots j}$ plus other tensors. For N=3 according to the short tensorial notation,
\begin{eqnarray}
\delta_{i i i j j j }= \delta_{i j}\delta_{i j} \delta_{i j} + \delta_{i i}\delta_{j j}\delta_{i j}.
\end{eqnarray}
To determine the number of components of this tensor, notice that for
$\delta_{i_1 i_2 i_3 j_1 j_2 j_3}$ once a pair is chosen, say $\delta_{i_1 j_1}$,  the previous N=2 is retrieved concerning the number of components. Since there are 5 ways to construct this first pair, the total number of components is 5 times 3, that is, 15 terms. The tensor $\delta_{i j}\delta_{i j} \delta_{i j}$ has 3! components, thus to know the number of terms in the tensor $\delta_{i i}\delta_{i j}\delta_{j j}$ we use the following argument. There are 3 components in $\delta_{i i}$, namely, $\delta_{i_1 i_2}$, $\delta_{i_1 i_3}$ and $\delta_{i_2 i_3}$ and similarly, 3 components in $\delta_{j j}$. Once fixed $\delta_{i i}$, and $\delta_{j j}$ the tensor $\delta_{i j}$ has only one possible component left. Therefore $\delta_{i i}\delta_{i j}\delta_{j j}$ has a total of 3 times 1 times 3, that is 9 components, and  the tensorial decomposition is expressed as $15 = 6 \oplus 9$.

Finally for N=4 the $i$ and $j$ short tensorial notation gives that,
\begin{eqnarray}
\delta_{i i i i j j j j }= \delta_{i j}\delta_{i j} \delta_{i j} \delta_{i j} + \delta_{i i}\delta_{j j}\delta_{i j}\delta_{i j} + \delta_{i i}\delta_{i i}\delta_{j j}\delta_{j j}
\end{eqnarray}
The same reasoning of the previous cases is used here, namely,  once a pair is fixed, say $\delta_{i_1 j_1}$, the number of terms of the remaining indices is provided by the previous N=3 case.
There are 7 ways to construct this first pair, thus the total number of terms is 7 times 15, that is, 105 terms. The tensor $\delta_{i j}\delta_{i j} \delta_{i j}\delta_{i j}$ has 4! terms,
The tensor $\delta_{i i}\delta_{j j}\delta_{i j}\delta_{i j}$ has 6 possible terms for $\delta_{i i}$, $\delta_{i_1 i_2}$, $\delta_{i_1 i_3}$, $\delta_{i_1 i_4}$, $\delta_{i_2 i_3}$, $\delta_{i_2 i_4}$   and $\delta_{i_3 i_4}$,  and the same applies for $\delta_{j j}$. Thus only 2 choices are left for $\delta_{i j}\delta_{i j}$, once the indices of $\delta_{i i}$ and $\delta_{j j}$ are fixed. Hence the total number of terms is 6 times 6 times 2, namely, 72 terms. For the tensor $\delta_{i i}\delta_{i i}\delta_{j j}\delta_{j j}$ the pairs $\delta_{i i}\delta_{i i}$ have only three terms each and so is for $\delta_{j j}\delta_{j j}$, such that the total is 3 times 3, namely, 9 terms.
The tensorial decomposition is expressed as $105 = 24 \oplus 72 \oplus 9$.\\

\section{Orthonormalization of polynomials to order N=4} \label{ortho}
The coefficients of the polynomials are determined here in terms of the integrals $I_N$. This is done for the first five polynomials by explicitly computing their inner products.
We refer to Eq.(\ref{omeg-feq}) by the short notation,
$$\left (\mathcal{P}_{(M)},\mathcal{P}_{(N)} \right )\equiv \int d^D \boldsymbol{\xi} \, \omega( \boldsymbol{\xi} )\mathcal{P}_{(M)}(\boldsymbol{\xi})\mathcal{P}_{(N)}(\boldsymbol{\xi},$$
used here for $M,N=0,\cdots 4$. The N$^{th}$ order polynomial is shortly referred as $\mathcal{P}_{(N)}$ such that the polynomials of Eqs.(\ref{H0}), (\ref{H1}), (\ref{H2}), (\ref{H3}), (\ref{H4}) are called $\mathcal{P}_{(0)}$, $\mathcal{P}_{(1)}$, $\mathcal{P}_{(2)}$, $\mathcal{P}_{(3)}$ and $\mathcal{P}_{(4)}$, respectively. The inner product between polynomials with distinct parity vanishes, $\left (\mathcal{P}_{(even)},\mathcal{P}_{(odd)} \right )=0$. Thus the only relevant orthonormality relations are among polynomials with the same parity (odd with odd and even with even).

The number of equations given by orthonormalization conditions must be equal to the number of free coefficients. Indeed this is the case, according to
Eqs.(\ref{H0}), (\ref{H1}), (\ref{H2}), (\ref{H3}), (\ref{H4}). The total number of coefficients is 14 ($c_K$ for $K=0,1,2,3,4$, ${c^\prime}_K$ for $K=2,3,4$, ${\bar c}_K$ for $K=2,3,4$, $d_4$, ${d^\prime}_4$ and ${\bar d}_4$).
Remarkably  14 equations arise from the 9 orthonormalization conditions $\left (\mathcal{P}_{(M)},\mathcal{P}_{(N)} \right )$, as seen below.
\\ \\
\noindent $\bullet \quad  \left (\mathcal{P}_{(0)},\mathcal{P}_{(0)} \right )$ \\ \\
The normalization of the N=0 polynomial is,
\begin{eqnarray}
c_0^2 \int d^D\boldsymbol{\xi}\omega(\xi)= 1,
\end{eqnarray}
which gives that,
\begin{eqnarray} \label{c_0}
c_0 = \pm\frac{1}{\sqrt{I_0}},
\end{eqnarray}
where Eq. (\ref{H0}) and the definition of $I_0$ in Eq.(\ref{i2n-cont}) have been used.
\\ \\
\noindent $\bullet \quad  \left (\mathcal{P}_{(1)},\mathcal{P}_{(1)} \right )$ \\ \\
\begin{eqnarray}
c_1^2\int d^D\boldsymbol{\xi}\; \omega(\xi)\xi_{i_1} \xi_{j_1} = \delta_{i_1 j_1}
\end{eqnarray}
The definition of $I_2$ in Eq.(\ref{i2n-cont}) is invoked to obtain that,
\begin{eqnarray} \label{c_1}
c_1 = \pm\frac{1}{\sqrt{I_2}}.
\end{eqnarray}
The  N=0 and N=1 polynomials are naturally orthogonal because they have distinct parity and to make them orthonormal is enough to normalize them which has been done above by determining the coefficients $c_0$ and $c_1$. The polynomial $\mathcal{P}_{(2)}$ has three coefficients, according to Eq.(\ref{H2}), and so, three equations are needed to determine them. These equations must arise from the tensorial structure of the D dimensional space.
\\ \\
\noindent $\bullet \quad  \left (\mathcal{P}_{(0)},\mathcal{P}_{(2)} \right )$ \\ \\
This conditions is equivalent to $c_0\left (1,\mathcal{P}_{(2)} \right )=0$ since $\mathcal{P}_{(0)}$ is a constant.
\begin{eqnarray}
c_0\int d^D\boldsymbol{\xi} \omega(\xi) \left [ c_2\,\xi_{i_1} \xi_{i_2} + f_2(\xi)\,\delta_{i_1 i_2}\right ] =0.
\label{orth-eq-2nd1}
\end{eqnarray}
Using the tensorial formulas of appendix~\ref{appendixb} and Eq.(\ref{i2n-cont}), it follows that,
\begin{eqnarray}
I_2c_2 + D I_2 \bar c_2 + I_0 c_2' = 0.
\label{orth-eq-2nd1b}
\end{eqnarray}
\\ \\
\noindent $\bullet \quad  \left (\mathcal{P}_{(2)},\mathcal{P}_{(2)} \right )$ \\ \\
Remarkably the normalization of the N=2 polynomial leads to multiple equations, in this case the following two equations.
\begin{eqnarray}
\int d^D\boldsymbol{\xi} \omega(\xi)\left [ c_2\,\xi_{i_1} \xi_{i_2} + f_2(\xi)\,\delta_{i_1 i_2}\right ]
\left [ c_2\,\xi_{j_1} \xi_{j_2} + f_2(\xi) \,\delta_{j_1 j_2}\right ]  =\delta_{i_1 j_1}\delta_{i_2 j_2} + \delta_{i_1 j_2}\delta_{i_2 j_1}.
\end{eqnarray}
This is because while the orthonormalization is associated to the tensor $ \delta_{i_1 j_1}\delta_{i_2 j_2} + \delta_{i_1 j_2}\delta_{i_2 j_1}$ the integration over $\xi_{i_1} \xi_{i_2} \xi_{j_1} \xi_{j_2}$ leads to the tensor $ \delta_{i_1 j_1}\delta_{i_2 j_2} + \delta_{i_1 j_2}\delta_{i_2 j_1}+\delta_{i_1 i_2}\delta_{j_1 j_2}$. This difference is responsible for the onset of more than one condition. Using the tensorial formulas of appendix~\ref{appendixb}, one obtains that:
\begin{eqnarray}
&& c_2^2 I_4 \big ( \delta_{i_1j_1}\delta_{i_2j_2} + \delta_{i_1j_2}\delta_{i_2j_1}+\delta_{i_1i_2}\delta_{j_1j_2} \big ) + \big [2(D+2)I_4c_2 \bar c_2 + 2I_2 c_2 c_2'+ \bar c_2^2 D(D+2)I_4 + 2DI_2\bar c_2 c_2' + c_2'^2I_0 \big ]\delta_{i_1i_2}\delta_{j_1j_2}\nonumber \\ && = \delta_{i_1j_1}\delta_{i_2j_2} +\delta_{i_1j_2}\delta_{i_2j_1}.
\end{eqnarray}
The independence of the two tensors leads to two independent equations, given by,
\begin{eqnarray}
 c_2^2 I_4 = 1  \:\: \mbox{and,}\:\:
 c^2_2I_4 + 2(D+2)I_4c_2 \bar c_2 + 2I_2 c_2 c_2'+ \bar c_2^2 D(D+2)I_4  + 2DI_2\bar c_2 c_2' + c_2'^2I_0 =0.
\label{orth-eq-2nd3}
\end{eqnarray}
The three equations are promptly solved and the coefficients $c_2$, ${\bar c}_2$ and ${c^\prime}_2$ are determined below.
\begin{eqnarray}\label{c_2}
c_2 = \pm \frac{1}{\sqrt{I_4}}.
\end{eqnarray}
Using Eq.\eqref{orth-eq-2nd1} to write $c_2'$, we have
\begin{eqnarray}\label{cp_2}
c_2' = -\frac{I_2}{I_0}c_2 - D\frac{I_2}{I_0}\bar c_2.
\end{eqnarray}
Substituting $c_1'$ in Eq.\eqref{orth-eq-2nd3},
\begin{eqnarray}
 \bar c_2^2 \big[ D(D+2) I_4 - D^2 \frac{I_2^2}{I_0} \big]  + \bar c_2c_2\big[ 2(D+2)I_4 - 2D\frac{I_2^2}{I_0} \big] + c_2^2 \big ( I_4-\frac{I_2^2}{I_0}\big) = 0,
\end{eqnarray}
which is a equation for $\bar c_2$. The solutions, are:
\begin{eqnarray}\label{cb_2}
 \bar c_2 = \frac{c_2}{D} (-1+\Delta_2), \: \mbox{where}\: \Delta _2 \equiv \pm \sqrt{\frac{2}{(D+2)-DJ_2}}, \; J_2 \equiv \frac{I_2^2 }{I_0 I_4}
\end{eqnarray}
The coefficient $\bar c_2$ must be a real number, and this is the case provided that  $(D+2)-DJ_2 \geq 0$. From Eq.\eqref{orth-eq-2nd1}, we calculate $c_2'$, to obtain that,
\begin{eqnarray}
c_2' = - c_2 \frac{I_2}{I_0} \Delta _2.
\end{eqnarray}
The orthogonalization of the first three polynomials has been concluded here. We proceed to the next order (N=3), and because of the increasing difficulty introduce a short notation for tensors, which is discussed in section~\ref{kron}. The N=3 polynomial  has three coefficients, similarly to the N=2 case, and so, three equations are needed to determine them.
\\ \\
\noindent $\bullet \quad  \left (\mathcal{P}_{(1)},\mathcal{P}_{(3)} \right)$ \\ \\
\begin{eqnarray}\label{orth-eq-3rd-i}
 c_1\int d^D\boldsymbol{\xi} \omega(\xi)\,\xi_{i_1} \: \big [ c_3 \,\xi_{j_1} \xi_{j_2}\xi_{j_3}  +  f_3(\xi)\, \big (\xi_{j_1} \delta_{j_2 j_3} + \xi_{j_2}\delta_{j_1 j_3} +\xi_{j_3} \delta_{j_1 j_2}\big) \big ]=0.
\end{eqnarray}
The integrals are calculated with the help of the tensorial formulas of appendix~\ref{appendixb}, such that the above expression becomes $\left [c_3I_4 + \bar c_3 I_4 (D+2) + c^{\prime}_3 I_2 \right ]\delta_{i j j j}=0 $.
The tensor $\delta_{i j j j}$ is short for $\delta_{i_1 j_1 j_2 j_3}$, according to the notation of section~\ref{kron}. Thus we obtain that,
\begin{eqnarray}\label{orth-eq-3rd-1}
c_3I_4 + \bar c_3 I_4 (D+2) + c^{\prime}_3 I_2 = 0.
\end{eqnarray}
\\ \\
\noindent $\bullet \quad  \left (\mathcal{P}_{(3)},\mathcal{P}_{(3)} \right )$ \\ \\
This normalization condition  gives the two other equations necessary to calculate the coefficients.
\begin{eqnarray}
&&\int d^D\boldsymbol{\xi} \omega(\xi) \big [ c_3 \,\xi_{i_1} \xi_{i_2}\xi_{i_3}  +f_3(\xi)\, \big (\xi_{i_1} \delta_{i_2 i_3} + \xi_{i_2}\delta_{i_1 i_3} + \xi_{i_3} \delta_{i_1 i_2}\big) \big ]\big [ c_3 \,\xi_{j_1} \xi_{j_2}\xi_{j_3}  +f_3(\xi)\, \big (\xi_{j_1} \delta_{j_2 j_3} + \xi_{j_2}\delta_{j_1 j_3}+\xi_{j_3} \delta_{j_1 j_2}\big) \big]  \nonumber \\
&& =
 \delta_{i_1 i_2 i_3 \vert j_1 j_2 j_3}
\label{orth-eq-3rd-2}
\end{eqnarray}
We stress the difference between the tensors $\delta_{i i i j j j }$ and  $\delta_{i i i \vert j j j }$ as the reason for multiple equations from a single normalization condition. Using the short notation of section~\ref{kron} the integral over the six vector components becomes,
$\int d^D\boldsymbol{\xi} \omega(\xi) \; \xi_{i}\xi_{i}\xi_{i}\xi_{j}\xi_{j}\xi_{j} = I_6 \delta_{i i i j j j }$, where all permutations of $i$'s and $j$'s are taken into account.
From the other side in this short notation, $\delta_{i i i \vert j j j} = \delta_{i j} \delta_{i j} \delta_{i j}$, as explained in section~\ref{kron}, and, as shown there, $\delta_{i i i j j j} = \delta_{i j} \delta_{i j} \delta_{i j}+\delta_{i i} \delta_{i j} \delta_{j j}$.
Thus the above integrals are computed with the aid of appendix \ref{appendixb}. Using this notation, Eq.\eqref{orth-eq-3rd-2} becomes,
\begin{eqnarray}
&&c_3^2I_6 \big(\delta_{ij}\delta_{ij}\delta_{ij} + \delta_{ii}\delta_{ij}\delta_{jj} \big ) + \big [2I_6 (D+4)c_3\bar c_3 + 2I_4c_3c^{\prime}_3 +  I_2 {c^{\prime}_3}^2+ 2I_4(D+2)\bar c_3 c^{\prime}_3+ I_6 (D+2) (D+4)\bar c_3^2 \big]\delta_{ii}\delta_{ij}\delta_{jj}\nonumber \\
&& =  \delta_{ij}\delta_{ij}\delta_{ij},
\end{eqnarray}
which gives two equations:
\begin{eqnarray}
&& c_3^2I_6 = 1 \:\:\mbox{and,}\:\: c_3^2I_6 + 2I_6 (D+4)c_3\bar c_3 + 2I_4c_3c^{\prime}_3 + I_2 {c^{\prime}_3}^2 + 2I_4(D+2)\bar c_3 c^{\prime}_3+ I_6 (D+2) (D+4)\bar c_3^2=0. \label{ortho-eq-3rd-b}
\end{eqnarray}
The solution of the first equation is,
\begin{eqnarray}\label{c_3}
c_3 = \pm \frac{1}{\sqrt{I_6}}.
\end{eqnarray}
Eq.\eqref{orth-eq-3rd-1} is used to eliminate $c^{\prime}_3$ from Eq.(\ref{ortho-eq-3rd-b}),
\begin{eqnarray} \label{eqbarc_3}
\bar c_3^2 (D+2) \left[ \left( I_6-\frac{I_4^2}{I_2}\right)(D+2) + 2I_6 \right] + 2 \left[\left( I_6-\frac{I_4^2}{I_2}\right)(D+2) + 2I_6 \right]c_3\bar c_3+ \left( I_6-\frac{I_4^2}{I_2}\right) c_3^2 =0.
\end{eqnarray}
This equation can be solved for $\bar c_3$.
\begin{eqnarray}\label{cb_3}
\bar c_3 = \frac{c_3}{D+2}(-1+\Delta _4),
\end{eqnarray}
and $c^{\prime}_3$ is calculated by Eq.\eqref{orth-eq-3rd-1}:
\begin{eqnarray}\label{cp_3}
 c^{\prime}_3 = - \frac{I_4}{I_2} \Delta _4 c_3 \:\: \mbox{where} \:\: \Delta_4 = \pm \sqrt{\frac{2}{(D+4) - J_4(D+2)}}, \, J_4 = \frac{I_4^2}{I_6 I_2}.
\end{eqnarray}
\\
The N=4 polynomial of Eq.(\ref{H4}) sets a new level of difficulty as  six equations must be obtained to calculate the six coefficients.
\\ \\
\noindent $\bullet \quad  \left (\mathcal{P}_{(0)},\mathcal{P}_{(4)} \right)$ \\ \\
The orthonormalization with the N=0 polynomial means that $c_0 \left (1,\mathcal{P}_{(4)} \right)$=0.
\begin{eqnarray}\label{orth-eq-4th-1}
&& c_0\int d^D \boldsymbol{\xi} \omega(\xi)
\; \big [c_4 \,\xi_{j_1} \xi_{j_2}\xi_{j_3} \xi_{j_4}+  f_4(\xi)\,\big (\xi_{j_1}\xi_{j_2}
\delta_{j_3 j_4} +   \xi_{j_1}\xi_{j_3}
\delta_{j_2 j_4}+\xi_{j_1}\xi_{j_4}
\delta_{j_2 j_3}+\xi_{j_2}\xi_{j_3} \delta_{j_1 j_4}+ \xi_{j_2}\xi_{j_4}
\delta_{j_1 j_3}\nonumber \\
&&+\xi_{j_3}\xi_{j_4}\delta_{j_1 j_2} \big )+g_4(\xi) \,\delta_{j_1 j_2 j_3 ij_4} \big ]=0
\end{eqnarray}
A single equation results from this integral since it can only be proportional to the tensor $\delta_{i i i i}$.
\begin{eqnarray}\label{orth-eq-4th-1}
 c_4I_4 + 2\big [c_4' I_2 + \bar c_4 I_4(D+2)\big ] + \nonumber  d_4I_0 + d_4' I_2 D + \bar d_4 I_4(D+2)D = 0.
\end{eqnarray}
\\ \\
\noindent $\bullet \quad  \left (\mathcal{P}_{(2)},\mathcal{P}_{(4)} \right)$ \\ \\
The integration of the N=2 with the N=4 polynomial gives that,
\begin{eqnarray}\label{orth-eq-4th-2}
&& \int d^D \boldsymbol{\xi} \omega(\xi) \big [c_2\,\xi_{i_1} \xi_{i_2} + f_2(\xi)\,\delta_{i_1 i_2}\big ]\big [c_4 \,\xi_{j_1} \xi_{j_2}\xi_{j_3} \xi_{j_4}+  f_4(\xi)\,\big (\xi_{j_1}\xi_{j_2}
\delta_{j_3 j_4} +\xi_{j_1}\xi_{j_3}
\delta_{j_2 j_4}+ \xi_{j_1}\xi_{j_4}\delta_{j_2 j_3}+\xi_{j_2}\xi_{j_3} \delta_{j_1 j_4} \nonumber \\
&&+ \xi_{j_2}\xi_{j_4}\delta_{j_1 j_3}+\xi_{j_3}\xi_{j_4}\delta_{j_1 j_2} \big )+g_4(\xi)\,\delta_{j_1 j_2 j_3 j_4} \big ]=0.
\end{eqnarray}
The  sixth order tensorial integral, $\int d^D \boldsymbol{\xi} \omega(\xi)\xi_{i}\xi_{i} \xi_{j} \xi_{j} \xi_{j} \xi_{j} = I_6 \delta_{i i j j j j}$ has 15 terms. This tensor can be decomposed as $\delta_{i i j j j j} = \delta_{i i}\delta_{j j j j}+\delta_{i j} \delta_{i j j j}$, namely, as a sum of two other tensors, which  have 3 and 12 terms, respectively. This decomposition can be formally expressed as $15=3\oplus 12$, as discussed in section~\ref{kron}.
We notice the presence of the tensor $\delta_{i_1 i_2 j_1 j_2}\delta_{j_3 j_4}+\delta_{i_1 i_2 j_1 j_3}\delta_{j_2 j_4}+\delta_{i_1 i_2 j_1 j_4}\delta_{j_2 j_3}+\delta_{i_1 i_2 j_2 j_3}\delta_{j_1 j_4}+\delta_{i_1 i_2 j_2 j_4}\delta_{j_1 j_3}+\delta_{i_1 i_2 j_3 j_4}\delta_{j_1 j_2}$, which has 18 terms, and is equal to $2\delta_{j j j j}\delta_{i i}+\delta_{i i i j}\delta_{i j}$ (2 times 3 plus 12). Therefore Eq.\eqref{orth-eq-4th-2} becomes,
\begin{eqnarray}
&&c_4c_2I_6 \big (\delta_{i i}\delta_{j j j j } + \delta_{i j}\delta_{i j j j} \big ) +  c_4 \big [c_2'I_4+ \bar c_2 I_6 (D+4)\big ]\delta_{i i}\delta_{j j j j} + c_2 \big [c_4'I_4 + \bar c_4 I_6(D+4)\big ]\big (2\delta_{i i}\delta_{j j j j}+ \delta_{i j}\delta_{i j j j}\big) + 2\big[c_4' c_2' I_2  \nonumber \\
&&+\big (\bar c_4 c_2'+ c_4' \bar c_2 \big )I_4 (D+2) + \bar c_4 \bar c_2 I_6 (D+4)(D+2)\big] \delta_{i i}\delta_{j j j j}  +c_2\big [d_4I_2+ d_4' I_4 (D+2) + I_6 \bar d_4 (D+4)(D+2)\big ]\delta_{i i}\delta_{j j j j }  \nonumber \\
&&+\big [d_4 c_2' I_0 + \big (d_4 \bar c_2 + d_4' c_2'\big )I_2D + \big (d_4' \bar c_2 + \bar d_4 c_2' \big )I_4(D+2)D  +\bar d_4 \bar c_2 I_6 (D+4)(D+2)D\big ] \delta_{i i}\delta_{j j j j} =0.
\end{eqnarray}
This lead to two equations, one proportional to $\delta_{ij}\delta_{ijjj}$,
\begin{eqnarray}\label{orth-eq-4th-3}
c_4I_6 + [c_4'I_4 + \bar c_4 I_6 (D+4)]=0,
\end{eqnarray}
and the other proportional to $\delta_{i i}\delta_{j j j j}$,
\begin{eqnarray}\label{orth-eq-4th-4}
&&c_4 c_2 I_6 + c_4 \big [ c_2' I_4 + \bar c_2 I_6 (D+4)\big ] + 2c_2\big [c_4'I_4 +
\bar c_4 I_6 (D+4)\big]+2\big[c_4' c_2' I_2+ \big (\bar c_4 c_2' + c_4' \bar c_2 \big )I_4 (D+2)\nonumber \\
&&+\bar c_4 \bar c_2 I_6 (D+4)(D+2)\big ] + c_2\big [d_4I_2 + d_4' I_4 (D+2) + \bar d_4 I_6 (D+4)(D+2)\big ]+ \big [d_4c_2'I_0 + \big (d_4 \bar c_2 + d_4' c_2' \big )I_2D\nonumber \\
&& +\big (d_4' \bar c_2 + \bar d_4 c_2' \big )I_4 (D+2)D + \bar d_4 \bar c_2 I_6 (D+4)(D+2)D\big ] = 0,
\end{eqnarray}
respectively.
\\ \\
\noindent $\bullet \quad  \left (\mathcal{P}_{(4)},\mathcal{P}_{(4)} \right)$ \\ \\
The normalization of the N=4 polynomial is given by,
\begin{eqnarray}\label{orth-4-by-4}
&&\int d^D \boldsymbol{\xi} \omega(\xi)\;\big[ c_4 \,\xi_{i_1} \xi_{i_2}\xi_{i_3} \xi_{i_4}+  f_4(\xi) \, \big (\xi_{i_1}\xi_{i_2}
\delta_{i_3 i_4} + \xi_{i_1}\xi_{i_3} \delta_{i_2 i_4} +\xi_{i_1}\xi_{i_4} \delta_{i_2 i_3}+\xi_{i_2}\xi_{i_3} \delta_{i_1 i_4}+ \xi_{i_2}\xi_{i_4}\delta_{i_1 i_3}+\xi_{i_3}\xi_{i_4}\delta_{i_1 i_2} \big )\nonumber\\
&& +g_4(\xi)\,\delta_{i_1 i_2 i_3 i_4}\big ] \big[ c_4 \,\xi_{j_1} \xi_{j_2}\xi_{j_3} \xi_{j_4}+  f_4(\xi) \, \big(\xi_{j_1}\xi_{j_2} \delta_{j_3 j_4} +\xi_{j_1}\xi_{j_3} \delta_{j_2 j_4} +\xi_{j_1}\xi_{j_4} \delta_{j_2 j_3}+\xi_{j_2}\xi_{j_3} \delta_{j_1 j_4}+ \xi_{j_2}\xi_{j_4}\delta_{j_1 j_3} \nonumber \\
&&+ \xi_{j_3}\xi_{j_4}\delta_{j_1 j_2} \big )+g_4(\xi)\,\delta_{j_1 j_2 j_3 j_4}\big] = \delta_{i_1 i_2 i_3 i_4 \vert j_1 j_2 j_3 j_4}.
\end{eqnarray}
Here we extensively use the short notation of section~\ref{kron},  where $\delta_{i i i i |j j j j } = \delta_{ij}\delta_{ij}\delta_{ij}\delta_{ij}$. There is the integral $\int d^D\boldsymbol{\xi} \omega(\xi) \; \xi_{i}\xi_{i}\xi_{i}\xi_{i}\xi_{j}\xi_{j}\xi_{j}\xi_{j} = I_8 \delta_{i i i i j j j j }$.
It holds that $\delta_{i i i i |j j j j} = \delta_{ij}\delta_{ij}\delta_{ij}\delta_{ij} + \delta_{ii}\delta_{jj}\delta_{ij}\delta_{ij} + \delta_{ii}\delta_{ii}\delta_{jj}\delta_{jj}$. There are 105 terms in  $\delta_{i i i i j j j j }$, 24 in  $\delta_{ij}\delta_{ij}\delta_{ij}\delta_{ij}$, 72 in $\delta_{ii}\delta_{jj}\delta_{ij}\delta_{ij}$, and 9 in $\delta_{ii}\delta_{ii}\delta_{jj}\delta_{jj}$.
We computed each of the  integrals individually and identify them by the following notation that uses the product of their coefficients $c_4$, $f_4$ and $g_4$. For instance the first one is
\begin{eqnarray}
 \int d^D \boldsymbol{\xi} \omega(\xi)\; c_4^2 [\cdots] = I_8 c_4^2 (\delta_{ij}\delta_{ij}\delta_{ij}\delta_{ij} +  \delta_{ii}\delta_{jj}\delta_{ij}\delta_{ij} + \delta_{ii}\delta_{ii}\delta_{jj}\delta_{jj})
\end{eqnarray}
The second term is,
\begin{eqnarray}
\int d^D \boldsymbol{\xi} \omega(\xi)\; c_4 f_4 [\cdots] = 2 [{c'}_{4}I_6+{\bar c}_4 I_8 (D+6) ] c_4  ( \delta_{ii}\delta_{jj}\delta_{ij}\delta_{ij} + 2 \delta_{ii}\delta_{ii}\delta_{jj}\delta_{jj}),
\end{eqnarray}
where we have used that $\delta_{i i i i j j} \delta_{j j}= \delta_{ii}\delta_{jj}\delta_{ij}\delta_{ij} + 2 \delta_{ii}\delta_{ii}\delta_{jj}\delta_{jj}$.
\begin{eqnarray}
 \int d^D \boldsymbol{\xi} \omega(\xi)\; c_4 g_4 [\cdots] = 2 [d_4 I_4 + d^{\prime}_4 I_6 (D+4) + \nonumber {\bar d}_4 I_8 (D+6) (D+4)] c_4  ( \delta_{ii}\delta_{ii}\delta_{jj}\delta_{jj}),
\end{eqnarray}
\begin{eqnarray}
&&\int d^D \boldsymbol{\xi} \omega(\xi)\; f_4 g_4 [\cdots] = 4 [c^{\prime}_4 d_4 I_2 + (c^{\prime}_4d^{\prime}_4+{\bar c}_4 d_4)I_4 (D+2)+({\bar c}_4 d^{\prime}_4+c^{\prime}_4 {\bar d}_4)I_6(D+4)(D+2)\nonumber \\
&&+{\bar c}_4 {\bar d}_4 I_8 (D+6) (D+4) (D+2)]
( \delta_{ii}\delta_{ii}\delta_{jj}\delta_{jj})
\end{eqnarray}
\begin{eqnarray}
&&\int d^D \boldsymbol{\xi} \omega(\xi)\; f_4^2 [\cdots] = [{c^{\prime}_4}^2 I_4+2c^{\prime}_4{\bar c}_4I_6(D+4)+{\bar c}_4^2I_8(D+6)(D+4)]
( 4 \delta_{ii}\delta_{ii}\delta_{jj}\delta_{jj}+ \delta_{ii}\delta_{jj}\delta_{ij}\delta_{ij})
\end{eqnarray}
\begin{eqnarray}
&&\int d^D \boldsymbol{\xi} \omega(\xi)\; g_4^2 [\cdots] = [{d_4}^2I_0+2d_4d^{\prime}_4 I_2 D + ({d^{\prime}_4}^2+2d_4 {\bar d}_4)I_4(D+2)D + 2 d^{\prime}_4{\bar d}_4 I_6 (D+4) (D+2) D +\nonumber \\
&&{{\bar d}_4}^2 I_8 (D+6)(D+4) (D+2) D ] ( \delta_{ii}\delta_{ii}\delta_{jj}\delta_{jj})
\end{eqnarray}
Next these integrals are introduced into Eq.(\ref{orth-4-by-4}) to obtain that,
\begin{eqnarray}
&&I_8(\delta_{ij}\delta_{ij}\delta_{ij}\delta_{ij} + \delta_{ii}\delta_{jj}\delta_{ij}\delta_{ij} + \delta_{ii}\delta_{ii}\delta_{jj}\delta_{jj}) c_4^2 + 2[c_4' I_6 + \bar c_4 I_8 (D+6)]c_4 (2\delta_{ii}\delta_{ii}\delta_{jj}\delta_{jj} + \delta_{ii}\delta_{jj}\delta_{ij}\delta_{ij}) + \nonumber\\
&&2[d_4I_4 + d_4'I_6 (D+4)+\bar d_4I_8 (D+6)(D+4)]\delta_{ii}\delta_{ii}\delta_{jj}\delta_{jj} + 4 [c_4' d_4 I_2 + (c_4' d_4' + \bar c_4 d_4)I_4 (D+2)+ \nonumber \\
&& (\bar c_4 d_4' + c_4' \bar d_4)I_6 (D+4)(D+2) + \bar c_4 \bar d_4 I_8 (D+6)(D+4)(D+2)]\delta_{ii}\delta_{ii}\delta_{jj}\delta_{jj} \nonumber \\
&&+ [c_4'^2I_4 + 2c_4' \bar c_4 I_6 (D+4)+ \bar c_4^2I_8(D+6)(D+4)] (4\delta_{ii}\delta_{ii}\delta_{jj}\delta_{jj} + \delta_{ii}\delta_{jj}\delta_{ij}\delta_{ij}) \nonumber\\
&&+[d_4^2 I_0 + 2d_4 d_4' I_2D + (d_4'^2 + 2d_4 \bar d_4)I_4 (D+2)D +2d_4' \bar d_4 I_6 (D+4)(D+2)D\nonumber \\
&&+ \bar d_4^2 I_8 (D+6)(D+4)(D+2)D] \delta_{ii}\delta_{ii}\delta_{jj}\delta_{jj} = \delta_{ij}\delta_{ij}\delta_{ij}\delta_{ij}.
\end{eqnarray}
The three remaining equations are the coefficients of the three independent tensors in the above equation.
\begin{eqnarray}\label{orth-eq-4th-6}
I_8 c_4^2 = 1,
\end{eqnarray}
\begin{eqnarray}\label{orth-eq-4th-7}
&&I_8 c_4^2 + 2[c_4'I_6 + \bar c_4 I_8(D+6)]c_4 + [c_4'^2 I_4 +  2c_4' \bar c_4 I_6 (D+4)+ \bar c_4^2 I_8 (D+6) (D+4)] = 0,
\end{eqnarray}
\begin{eqnarray}\label{orth-eq-4th-8}
&&I_8 c_4^2 + 4c_4[c_4' I_6 + \bar c_4 I_8(D+6)] + 2c_4[d_4 I_4 + d_4'I_6(D+4)+ \bar d_4 I_8 (D+6)(D+4)] + 4[c_4' d_4 I_2 + \nonumber\\
&&(c_4'd_4' + \bar c_4 d_4)I_4(D+2)+ (\bar c_4 d_4' + c_4' \bar d_4)I_6 (D+4)(D+2) + \bar c_4 \bar d_4 I_8 (D+6)(D+4)(D+2)]+ 4[c_4'^2I_4 \nonumber \\
&&+ 2c_4'\bar c_4 I_6 (D+4)  +\bar c_4^2 I_8 (D+6)(D+4)] + [d_4^2I_0 + 2d_4d_4'I_2D + (d_4'^2+ 2d_4 \bar d_4)I_4 (D+2)D \nonumber\\
&&+ 2d_4' \bar d_4 I_6 (D+4)(D+2)D + \bar d_4^2 I_8 (D+6)(D+4)(D+2)D] = 0.
\end{eqnarray}
The six equations given by the Eqs.\eqref{orth-eq-4th-1}, \eqref{orth-eq-4th-3}, \eqref{orth-eq-4th-4}, \eqref{orth-eq-4th-6}, \eqref{orth-eq-4th-7} and \eqref{orth-eq-4th-8} can be solved to obtain the six coefficients.
Nevertheless notice that coefficients $c_4$, $\bar c_4$ and $c^{\prime}_4$ are determined from a sub set of equations, namely, Eqs.(\ref{orth-eq-4th-3}), (\ref{orth-eq-4th-6}),
and (\ref{orth-eq-4th-7}). From them it follows that,
\begin{eqnarray}
&& c_4 = \pm\frac{1}{\sqrt{I_8}},\, \mbox{and},\label{c_4} \\
&&\bar c_4^2 (D+4) \big[ I_8 (D+6)-\frac{I_6^2}{I_4}(D+4)\big] + 2 \big[ I_8 ((D+6)-\frac{I_6^2}{I_4}(D+4) + \big]c_4\bar c_4+ \big( I_8-\frac{I_6^2}{I_4}\big) c_4^2 =0. \label{eqbarc_4}
\end{eqnarray}
Hence one obtains that,
\begin{eqnarray}
&& \bar c_4 = c_4\frac{(-1 + \Delta _6)}{D+4}, \mbox{and} \label{cb_4}\\
&& c_4' = - \frac{I_6}{I_4}\Delta _6 c_4. \label{cp_4}
\end{eqnarray}
The remaining coefficients are given by,
\begin{eqnarray}\label{db_4}
\bar d_4 = \frac{\delta _2}{\delta _4D(D+2)} d_4 + \frac{c_4[D-2(D+2)\Delta_6]}{D(D+2)(D+4)},
\end{eqnarray}
\begin{eqnarray}\label{dp_4}
d_4' = -\frac{d_4}{D}\left( \frac{I_0}{I_2} + \frac{I_4}{I_2}\frac{\delta _2}{\delta _4} \right) + \frac{2I_6\Delta_6 c_4}{I_4D},
\end{eqnarray}
and,
\begin{eqnarray}\label{d_4}
d_4 = \pm  \sqrt{\frac{8\delta _4^2 I_4}{\delta _2}}\frac{1}{\sqrt{\delta _2 \delta _6(D+4) - \delta _4^2D}},
\end{eqnarray}
where
\begin{eqnarray}
\Delta _6 = \pm \sqrt{\frac{2}{(D+6)- J_6(D+4)}},\:\:\: J_6 = \frac{I_6^2}{I_8I_4},
\end{eqnarray}
\begin{eqnarray}
&& \delta _2 = I_0 I_4(D+2) - I_2^2D,\\
&& \delta _4 = I_2I_6(D+4) - I_4^2(D+2),\\
&& \delta _6 = I_4 I_8 (D+6) - I_6^2 (D+4).
\end{eqnarray}
Notice that $\delta_{2K} =2I_{2K+2}I_{2K-2}/\Delta_{2K}^2$.\\

Hence we have determined all the coefficients in Eqs.(\ref{c_0}), (\ref{c_1}), (\ref{c_2}), (\ref{cp_2}), (\ref{cb_2}), (\ref{c_3}), (\ref{cb_3}), (\ref{cp_3}), (\ref{c_4}), (\ref{cb_4}), (\ref{cp_4}), (\ref{db_4}), (\ref{dp_4}), and (\ref{d_4}).
Some of the coefficients can be summarized in  simple formulas for all polynomials, using a general notation:
\begin{eqnarray}
&& c_K= \frac{1}{\sqrt{I_{2K}}},\; \mbox{for  $K=0,1,2,3,4$}, \nonumber \\
&& {c^\prime}_K=-c_K \frac{I_{2K-2}}{I_{2K-4}}\Delta_{2K-2},\; \mbox{for  $K=2,3,4$},  \nonumber \\
&& {\bar c}_K=c_K \frac{\left (-1+\Delta_{2K-2}\right)}{D+2K-4},\; \mbox{for  $K=2,3,4$}, \nonumber \\
&& \Delta_{2K}= \sqrt{\frac{2}{\big(D+2K\big)-J_{2K}\big(D+2K-2\big)}} \nonumber \\
&& J_{2K}= \frac{I_{2K}^2}{I_{2K+2}I_{2K-2}} \nonumber
\end{eqnarray}
Notice that we have chosen the positive solutions for the square roots, but the negative ones would also lead to orthogonal polynomials. We have completed the orthonormalization of the first five polynomials (N=0,1,2,3,4). The procedure can be applied for higher polynomials, although it becomes increasingly laborious.

\section{D-dimensional polynomials for specific weights}\label{secweight}
In this section we obtain the D-dimensional polynomials for some specific weights and from them retrieve some well known D=1 polynomials. We also obtain the $I_N$ functions associated to some new weights.

\subsection{D-dimensional Hermite polynomials} \label{dherm}
We show that the D-dimensional Hermite polynomials are straightforwardly retrieved from the above polynomials for the gaussian weight,
\begin{eqnarray}
\omega(\xi) = \frac{1}{(2\pi)^{D/2}}e^{-\frac{\xi^2}{2}},\, \mbox{and} \; \xi_{max}=\infty.
\end{eqnarray}
To obtain the integrals $I_{2N}$ of Eq.(\ref{i2n-cont2}), we note that,
\begin{eqnarray}\label{i2n-cont2}
\int_{0}^{\infty} d\xi\, \omega(\xi)\,\xi^{2N+D-1}= \frac{2^{N-1}}{\pi^{\frac{D}{2}}}\Gamma\big(N+\frac{D}{2}\big).
\end{eqnarray}
Then it follows from Eq.(\ref{i2n-cont2}) that,
\begin{eqnarray}\label{i2n-hermite}
I_{2N}=1.
\end{eqnarray}
In this limit $c_K=1$, $\bar c_K=0$  $c^{\prime}_K=-1$,  $d_4=1$, $\bar d_4=0$ and $d^{\prime}_4=0$, and the polynomials of Eqs.(\ref{H0}), (\ref{H1}), (\ref{H2}), (\ref{H3}), (\ref{H4}) become,
\begin{eqnarray}
&&\mathcal{P}_0(\boldsymbol{\xi}) = 1,\label{Her0}
\end{eqnarray}
\begin{eqnarray}
&&\mathcal{P}_{i_1}(\boldsymbol{\xi})=\,\xi_{i_1},\label{Her1}
\end{eqnarray}
\begin{eqnarray}
&&\mathcal{P}_{i_1 i_2}(\boldsymbol{\xi})=\,\xi_{i_1} \xi_{i_2} - \delta_{i_1 i_2}\label{Her2}
\end{eqnarray}
\begin{eqnarray}
\mathcal{P}_{i_1 i_2 i_3}(\boldsymbol{\xi})= \,\xi_{i_1} \xi_{i_2}\xi_{i_3}  -\big (\xi_{i_1} \delta_{i_2 i_3} + \xi_{i_2}\delta_{i_1 i_3} +  \xi_{i_3} \delta_{i_1 i_2}\big)\label{Her3}
\end{eqnarray}
and,
\begin{eqnarray}
&&\mathcal{P}_{i_1 i_2 i_3 i_4}(\boldsymbol{\xi})= \,\xi_{i_1} \xi_{i_2}\xi_{i_3} \xi_{i_4}-\,\big(\xi_{i_1}\xi_{i_2}
\delta_{i_3 i_4} +\xi_{i_1}\xi_{i_3}
\delta_{i_2 i_4} \nonumber \\
&&+\xi_{i_1}\xi_{i_4}
\delta_{i_2 i_3} +\xi_{i_2}\xi_{i_3} \delta_{i_1 i_4}+ \xi_{i_2}\xi_{i_4}
\delta_{i_1 i_3} +\xi_{i_3}\xi_{i_4}\delta_{i_1 i_2} \big )+\,\delta_{i_1 i_2 i_3 i_4}\label{Her4}
\end{eqnarray}
We notice that the tensorial basis that spans the new generalized polynomials contains the basis that spans the Hermite polynomials but not vice-versa. The D-dimensional Hermite polynomials $\mathcal{P}_{i_1\cdots i_N}$ are symmetric tensors in the  indices $i_1\cdots i_N$ spanned over the basis formed by the tensors,
$$T_{i_1\cdots i_N}\equiv\xi_{i_1}\cdot \xi_{i_2}\cdots \xi_{i_P}\cdot \delta_{i_{P+1},i_{P+2}}\cdot\delta_{i_{P+3},i_{P+4}}\cdots\delta_{i_{N-1},i_{N}}.$$
This basis is not large enough to span the new generalized polynomials,
$\mathcal{P}_{i_1\cdots i_N}$, which demand a larger basis formed by the tensors
$$T_{i_1\cdots i_N}\equiv F\big( \xi\big) \xi_{i_1}\cdot \xi_{i_2}\cdots \xi_{i_P}\cdot \delta_{i_{P+1},i_{P+2}}\cdot \delta_{i_{P+3},i_{P+4}}\cdots\delta_{i_{N-1},i_{N}},$$
whose scalar functions $F\big( \xi \big)$ are polynomials in powers of the vector modulus, $1$, $\xi^2$, $\xi^4$, ...,$\xi^{2k}$.
\\

\subsection{D-dimensional Legendre polynomials}\label{dlegen}
We define D-dimensional Legendre polynomials satisfying the general orthonormal relation of Eq.(\ref{omeg-feq}) with  $\xi_{max} = 1$ and the following weight:
\begin{eqnarray}
\omega(\xi) = 1, \:\: \mbox{and} \:\: \xi_{max} = 1.
\end{eqnarray}
From this it follows that,
\begin{eqnarray}
I_{2N} = \frac{\pi^{D/2}}{2^{N-1}\Gamma(N+D/2)} \int_0^1 d\xi \xi^{2N + D -1}= \frac{2^{1-N} \pi ^{D/2}}{(D+2N) \Gamma\left[\frac{D+2N}{2}\right]},
\end{eqnarray}
Using the expressions  corresponding coefficients up to fourth order
\begin{eqnarray}
c_N=\frac{1}{\sqrt{\frac{2^{-N} \pi ^{D/2}}{\Gamma\left[1+\frac{D}{2}+N\right]}}},
\end{eqnarray}
\begin{eqnarray}
\bar c_{2}=\frac{-2+ \sqrt{2(2+D)}}{D \sqrt{\frac{\pi ^{D/2}}{\Gamma\left[3+\frac{D}{2}\right]}}},\:\;
\bar c_{3}= \frac{ -2\sqrt{2}+2\sqrt{4+D}}{(2+D) \sqrt{\frac{\pi ^{D/2}}{\Gamma\left[4+\frac{D}{2}\right]}}},\:\:
\bar c_4 =\frac{ -4 + 2\sqrt{2(6+D)}}{(4+D) \sqrt{\frac{\pi ^{D/2}}{\Gamma\left[5+\frac{D}{2}\right]}}},
\end{eqnarray}
\begin{eqnarray}
c^{\prime}_{2}=-\sqrt{\frac{2}{(2+D )\frac{\pi ^{D/2}}{\Gamma\left[3+\frac{D}{2}\right]}}}, \:\:
c^{\prime}_{3}=-\sqrt{\frac{4}{(4+D) \frac{\pi ^{D/2}}{\Gamma\left[4+\frac{D}{2}\right]}}},\:\:
c^{\prime}_4=-\sqrt{\frac{8}{(6+D) \frac{\pi ^{D/2}}{\Gamma\left[5+\frac{D}{2}\right]}}},
\end{eqnarray}
\begin{eqnarray}
d_4=\sqrt{\frac{ \frac{16(6+D)\pi ^{3 D/2}}{(4+D) \Gamma\left[3+\frac{D}{2}\right] \Gamma\left[4+\frac{D}{2}\right]^2}}{\pi
^{2 D} \left(\frac{(2+D) (96+D (8+D) (24+D (8+D)))}{(8+D) \Gamma\left[3+\frac{D}{2}\right]^4}-\frac{2 (4+D) (4+D (8+D))}{\Gamma\left[2+\frac{D}{2}\right]^3
\Gamma\left[5+\frac{D}{2}\right]}\right)}},
\end{eqnarray}
\begin{eqnarray}
d^{\prime}_4=\frac{\left( 2 D (6+D)^{3/2} \Gamma\left[1+\frac{D}{2}\right] \Gamma\left[3+\frac{D}{2}\right]^2 - 8\sqrt{6+D} \Gamma\left[3+\frac{D}{2}\right]^3    \right)d_4 +  \left( \frac{8 \sqrt{2} \Gamma\left[3+\frac{D}{2}\right]^3 }{D+2}    -4 \sqrt{2} \Gamma\left[2+\frac{D}{2}\right]^2 \Gamma\left[4+\frac{D}{2}\right] \right)c_4    }{D (4+D) \sqrt{6+D} \left(-(6+D) \Gamma\left[2+\frac{D}{2}\right]^3+2 \Gamma\left[1+\frac{D}{2}\right]
\Gamma\left[3+\frac{D}{2}\right]^2\right)},
\end{eqnarray}
\begin{eqnarray}
\bar d_4&=&\frac{c_4 \left(D-2  \sqrt{2(6+D)}- D \sqrt{2(6+D)}\right)+(4+D)^2 (6+D) d_4}{D (2+D) (4+D)}.
\end{eqnarray}
These coefficients provides orthogonal polynomials for any dimension and, as we will see in subsection~\ref{projection-1d},
they give the well known D=1 Legendre polynomials for the particular case with $D=1$.

\subsection{D-dimensional Chebyshev polynomials of first kind}\label{dcheb1}

Following the same procedure used for the Legendre polynomials, we define the D-dimensional Chebyshev polynomials
of first kind with the weight $\omega(\xi)$ below and integrated under a sphere of radius $\xi_{max}$,
\begin{eqnarray}
\omega(\xi) = \frac{1}{1-\xi^2}, \:\: \mbox{and} \:\: \xi_{max} = 1  ,
\end{eqnarray}
with the corresponding integral
\begin{eqnarray}
I_{2N} = \frac{\pi^{D/2}}{2^{N-1}\Gamma(N+D/2)} \int_0^1 d\xi \frac{1}{1-\xi^2} \xi^{2N + D -1}
= \frac{2^{-N} \pi ^{\frac{1+D}{2}}}{\Gamma\left[\frac{1}{2} (1+D+2N)\right]},
\end{eqnarray}
and coefficients up to fourth order
\begin{eqnarray}
c_N=\frac{1}{\sqrt{\frac{2^{-N} \pi ^{\frac{1+D}{2}}}{\Gamma\left[\frac{1+D}{2}+N\right]}}},
\end{eqnarray}
\begin{eqnarray}
\bar c_{2}=\frac{2 \left(-1+\sqrt{1+D}\right)}{D \sqrt{\frac{\pi ^{\frac{1+D}{2}}}{\Gamma\left[\frac{5+D}{2}\right]}}},\:\:
\bar c_{3}=\frac{2 \sqrt{2} \left(-1+\sqrt{3+D}\right)}{(2+D) \sqrt{\frac{\pi ^{\frac{1+D}{2}}}{\Gamma\left[\frac{7+D}{2}\right]}}},\:\: \bar c_4 =\frac{4 \left(-1+\sqrt{5+D}\right)}{(4+D) \sqrt{\frac{\pi ^{\frac{1+D}{2}}}{\Gamma\left[\frac{9+D}{2}\right]}}},
\end{eqnarray}
\begin{eqnarray}
c^{\prime}_{2}=-\frac{2}{ \sqrt{\frac{(1+D)\pi ^{\frac{1+D}{2}}}{\Gamma\left[\frac{5+D}{2}\right]}}},\:\:
c^{\prime}_{3}=-\frac{2 \sqrt{2}}{ \sqrt{\frac{(3+D)\pi ^{\frac{1+D}{2}}}{\Gamma\left[\frac{7+D}{2}\right]}}}, \:\: c_4'=-\frac{4}{ \sqrt{\frac{(5+D)\pi ^{\frac{1+D}{2}}}{\Gamma\left[\frac{9+D}{2}\right]}}}
\end{eqnarray}
\begin{eqnarray}
d_4=\frac{8 \sqrt{\frac{\pi ^{\frac{3 (1+D)}{2}}}{(3+D) (5+D)^2 \Gamma\left[\frac{5+D}{2}\right]^3}}}{\sqrt{\pi ^{2+2 D} \left(-\frac{D
(2+D)^2}{\Gamma\left[\frac{5+D}{2}\right]^4}+\frac{(4+D) (7+D) (4+D (7+D) (8+D (7+D)))}{4 (3+D) \Gamma\left[\frac{3+D}{2}\right]^2 \Gamma\left[\frac{9+D}{2}\right]^2}\right)}},
\end{eqnarray}
\begin{eqnarray}
d^{\prime}_4 &=&  \frac{2 \left(c_4 \sqrt{5+D}-d_4 (3+D) (5+D) \right) \Gamma\left[\frac{3+D}{2}\right] \Gamma\left[\frac{5+D}{2}\right]^2}{D
(4+D) (5+D) \Gamma\left[\frac{5+D}{2}\right]^3-2 D (2+D) \Gamma\left[\frac{3+D}{2}\right] \Gamma\left[\frac{7+D}{2}\right]^2},
\end{eqnarray}
\begin{eqnarray}
&&\bar d_4 = \frac{1}{4 (2+D)}\left(\frac{4 c_4 \left(D-4 \sqrt{5+D}-2 D \sqrt{5+D}\right)}{D (4+D)}+\frac{8 d_4 (5+D)  \Gamma\left[\frac{5+D}{2}\right]^2}{D
(4+D) \Gamma\left[\frac{5+D}{2}\right]^2-D (2+D) \Gamma\left[\frac{3+D}{2}\right] \Gamma\left[\frac{7+D}{2}\right]}\right).
\end{eqnarray}
The polynomials with the coefficients above are orthogonal for any dimensions and, as we will see in subsection~\ref{projection-1d} for $D=1$ they give the well known
D=1 Chebyshev polynomials of first kind.

\subsection{D-dimensional Chebyshev polynomials of second kind}\label{dcheb2}

For the D-dimensional Chebyshev polynomials of second kind, we define the following weight function and maximum radius
\begin{eqnarray}
\omega(\xi) = \sqrt{1-\xi^2}, \:\: \mbox{and} \:\: \xi_{max} = 1
\end{eqnarray}
which give the integral and coefficients below
\begin{eqnarray}
I_{2N} = \frac{\pi^{D/2}}{2^{N-1}\Gamma(N+D/2)} \int_0^1 d\xi \sqrt{1-\xi^2} \xi^{2N + D -1}
= \frac{2^{-1-N} \pi ^{\frac{1+D}{2}}}{\Gamma\left[\frac{1}{2} (3+D+2N)\right]},
\end{eqnarray}
\begin{eqnarray}
c_N=\frac{1}{\sqrt{\frac{2^{-1-N} \pi ^{\frac{1+D}{2}}}{\Gamma\left[\frac{3+D}{2}+N\right]}}},
\end{eqnarray}
\begin{eqnarray}
\bar c_{2}=\frac{2 \sqrt{2} \left(-3+ \sqrt{3(3+D)}\right)}{3 D \sqrt{\frac{\pi ^{\frac{1+D}{2}}}{\Gamma\left[\frac{7+D}{2}\right]}}},\:\:
\bar c_{3}=\frac{-4+4 \sqrt{5+D}/\sqrt{3}}{(2+D) \sqrt{\frac{\pi ^{\frac{1+D}{2}}}{\Gamma\left[\frac{9+D}{2}\right]}}},\:\:
\bar c_4 =\frac{4 \sqrt{2} \left(-3+ \sqrt{3(7+D)}\right)}{3 (4+D) \sqrt{\frac{\pi ^{\frac{1+D}{2}}}{\Gamma\left[\frac{11+D}{2}\right]}}},
\end{eqnarray}
\begin{eqnarray}
c^{\prime}_{2}=-\sqrt{\frac{8}{ \frac{3(3+D)\pi ^{\frac{1+D}{2}}}{\Gamma\left[\frac{7+D}{2}\right]}}},\:\:
c^{\prime}_{3}=-\sqrt{\frac{16}{ \frac{3(5+D)\pi ^{\frac{1+D}{2}}}{\Gamma\left[\frac{9+D}{2}\right]}}},\:\:
c^{\prime}_4=-\sqrt{\frac{32}{ \frac{3(7+D)\pi ^{\frac{1+D}{2}}}{\Gamma\left[\frac{11+D}{2}\right]}}}
\end{eqnarray}
\begin{eqnarray}
d_4=\frac{4 \sqrt{\frac{3(7+D) \pi ^{\frac{3 (1+D)}{2}}}{(5+D) \Gamma\left[\frac{9+D}{2}\right]^3}}}{\sqrt{\pi ^{2+2 D} \left(-\frac{D
(2+D)^2}{\Gamma\left[\frac{7+D}{2}\right]^4}+\frac{(4+D) (9+D) (36+D (1+D) (8+D) (9+D))}{4 (5+D) \Gamma\left[\frac{5+D}{2}\right]^2 \Gamma\left[\frac{11+D}{2}\right]^2}\right)}}
\end{eqnarray}
\begin{eqnarray}
d_4^\prime = \frac{2 \left(c_4  \sqrt{3(7+D)}-3 d_4(5+D) (7+D) \right) \Gamma\left[\frac{5+D}{2}\right] \Gamma\left[\frac{7+D}{2}\right]^2}{D
(4+D) (7+D) \Gamma\left[\frac{7+D}{2}\right]^3-2 D (2+D) \Gamma\left[\frac{5+D}{2}\right] \Gamma\left[\frac{9+D}{2}\right]^2}
\end{eqnarray}
\begin{eqnarray}
&&\bar d_4 = \frac{1}{4
(2+D)}\left(-\frac{4 c_4 \left(4 \sqrt{3(7+D)}+D \left(-3+2 \sqrt{3(7+D)}\right)\right)}{3 D (4+D)}+\frac{24 d_4 (7+D) \Gamma\left[\frac{7+D}{2}\right]^2}{D (4+D) \Gamma\left[\frac{7+D}{2}\right]^2-D (2+D) \Gamma\left[\frac{5+D}{2}\right] \Gamma\left[\frac{9+D}{2}\right]}\right) \\ \nonumber
&&
\end{eqnarray}
The corresponding polynomials are orthogonal for any dimension and, as we will see in subsection~\ref{projection-1d}, give the well known D=1 Chebyshev polynomials of second kind.

\subsection{Projection of Hermite, Legendre and Chebyshev polynomials into  D=1 dimension}\label{proj1d}
\label{projection-1d}
To obtain the projection of such polynomials in D=1 dimension, it suffices to drop the index of the vector, $\xi_{i_1} \rightarrow \xi $, and to take that $\delta_{i_1 i_2} \rightarrow 1$ since there is only one index, and so, $i_1=i_2=1$.
Thus the tensors based on the Kronecker's delta function have each of its terms equal to one, for instance, $\delta_{i_1 j_1}\cdots \delta_{i_Nj_N}=1$, and so, $\delta_{i_1\cdots i_N\vert j_1\cdots j_N}=N\, \!!$ and $\delta_{i_1\cdots i_N\, j_1\cdots j_N}= (2N-1)\,\!!/[2^{N-1}(N-1)\,\!!]$, according to section~\ref{kron}.
Hence the orthonormality condition of Eq.(\ref{omeg-feq}) becomes,
\begin{flalign}
\int d^D \boldsymbol{\xi} \, \omega( \boldsymbol{\xi} )\mathcal{P}_{N}( \boldsymbol{\xi})\mathcal{P}_{M}(\boldsymbol{\xi})= N\, \!!
\delta_{\scriptscriptstyle {N M}} \label{omeg-feq1d}
\end{flalign}
Notice that this is not the standard normalization employed in the definition of most of D=1 orthonormal polynomials.
\subsubsection{\bf  Hermite polynomials}
The Hermite polynomials obtained via dimensional reduction from Eqs.(\ref{Her0}), (\ref{Her1}), (\ref{Her2}), (\ref{Her3}) and (\ref{Her4}) are given by,
\begin{flalign}
\mathcal{P}_0(\boldsymbol{\xi}) = 1, \:\:
\mathcal{P}_{1}(\boldsymbol{\xi})=\xi,\:\:
\mathcal{P}_{2}(\boldsymbol{\xi})=\xi^2 - 1,\:\:
\mathcal{P}_{3}(\boldsymbol{\xi})=\xi^3  -3\xi, \:\: \mbox{and}, \:\:
\mathcal{P}_{4}(\boldsymbol{\xi})= \xi^4-6\xi^2+3.
\end{flalign}
The known D=1 Hermite  polynomials, defined with the normalization of probability theory~\cite{gradshteyn14}, satisfy the following orthonormality condition, $$\int _{-\infty }^{\infty }{\mathit {He}}_{m}(x){\mathit {He}}_{n}(x)\,e^{-{\frac {x^{2}}{2}}}\,\mathrm {d} x={\sqrt {2\pi }}n!\delta _{nm}.$$ The first five ones  are
${{\mathit {He}}_{0}(x)=1}$, ${{\mathit {He}}_{1}(x)=x}$, ${ {\mathit {He}}_{2}(x)=x^{2}-1}$,
${{\mathit {He}}_{3}(x)=x^{3}-3x}$, and ${{\mathit {He}}_{4}(x)=x^{4}-6x^{2}+3}$ and they coincide exactly with the present polynomials.
\subsubsection{\bf Legendre polynomials }
It follows by taking $D=1$ that,
\begin{flalign}
\mathcal{P}_{0}(\boldsymbol{\xi}) =   \frac{1}{\sqrt{2}},\:\:
\mathcal{P}_{1}(\boldsymbol{\xi}) =  \frac{3}{2} \xi, \:\:
\mathcal{P}_{2}(\boldsymbol{\xi}) =  \frac{\sqrt{5}}{2}(3\xi^2 -1),
\end{flalign}
\begin{flalign}
\mathcal{P}_{3}(\boldsymbol{\xi}) =  \frac{\sqrt{21}}{2}(5\xi^2 - 3), \, \mbox{and} \:\:
\mathcal{P}_{4}(\boldsymbol{\xi}) = \frac{3 \sqrt{3}}{4} ( 35 \xi^4 - 30 \xi^2 + 3).
\end{flalign}
The Legendre  polynomials~\cite{gradshteyn14} satisfy the following orthonormality condition, $$\int _{-1 }^{1 }{\mathit {P}}_{m}(x){\mathit {P}}_{n}(x)\,\mathrm {d} x=\frac{2}{2n+1}\delta _{nm}.$$ The first five ones  are
${{\mathit {P}}_{0}(x)=1}$, ${{\mathit {P}}_{1}(x)=x}$, ${ {\mathit {P}}_{2}(x)=(3x^{2}-1)/2}$,
${{\mathit {P}}_{3}(x)=(5x^{3}-3x)/2}$, and ${{\mathit {P}}_{4}(x)=(35x^{4}-30x^{2}+3)/8}$. Thus apart from the normalization they coincide with the present polynomials.
\subsubsection{\bf Chebyshev polynomials of the first kind}
\begin{flalign}
\mathcal{P}_{0}(\boldsymbol{\xi}) =  \frac{1}{\sqrt{\pi}},\:\:
\mathcal{P}_{1}(\boldsymbol{\xi}) =  \sqrt{\frac{2}{\pi}}\xi,\:\:
\mathcal{P}_{2}(\boldsymbol{\xi}) =  \frac{2}{\sqrt{\pi}}(2\xi^2 -1),
\end{flalign}
\begin{flalign}
\mathcal{P}_{3}(\boldsymbol{\xi}) = 2\sqrt{\frac{3}{\pi}} (4\xi^3 - 3\xi),\, \mbox{and} \:\:
\mathcal{P}_{4}(\boldsymbol{\xi}) = 4 \sqrt{\frac{3}{\pi }} \left( 8 \xi^4 - 8 \xi^2 + 1\right).
\end{flalign}
The  Chebyshev  polynomials of the first kind~\cite{gradshteyn14} satisfy the following orthonormality condition,
$$ {\displaystyle \int _{-1}^{1}{\mathit{T}}_{n}(x){\mathit{T}}_{m}(x)\,{\frac {dx}{\sqrt {1-x^{2}}}}={\begin{cases}0&n\neq m\\\pi &n=m=0\\{\frac {\pi }{2}}&n=m\neq 0\end{cases}}}.$$
The first five ones  are
${{\mathit {T}}_{0}(x)=1}$, ${{\mathit {T}}_{1}(x)=x}$, ${ {\mathit {T}}_{2}(x)=2x^{2}-1}$,
${{\mathit {T}}_{3}(x)=4x^{3}-3x}$, and ${{\mathit {T}}_{4}(x)=8x^{4}-8x^{2}+1}$. Thus apart from the normalization they coincide with the present polynomials.
\subsubsection{\bf Chebyshev polynomials of the second kind}
\begin{flalign}
\mathcal{P}_{0}(\boldsymbol{\xi}) =  \sqrt{\frac{2}{\pi}}, \:\:
\mathcal{P}_{1}(\boldsymbol{\xi}) =  2 \sqrt{2}{\pi} \xi,\:\:
\mathcal{P}_{2}(\boldsymbol{\xi}) =  \frac{2}{\sqrt{\pi}}(4\xi^2 -1),
\end{flalign}
\begin{flalign}
\mathcal{P}_{3}(\boldsymbol{\xi}) = 8 \sqrt{\frac{3}{\pi}} (2\xi^3 - \xi),\:\: \mbox{and},\:\:
\mathcal{P}_{4}(\boldsymbol{\xi}) = 4 \sqrt{\frac{3}{\pi }} \left(16 \xi^4 -12 \xi^2 + 1\right).
\end{flalign}
The  Chebyshev  polynomials of the second kind~\cite{gradshteyn14} satisfy the following orthonormality condition,
$$ {\displaystyle \int _{-1}^{1}{\mathit {U}}_{n}(x){\mathit {U}}_{m}(x)\,{\frac {dx}{\sqrt {1-x^{2}}}}={\begin{cases}0&n\neq m\\{\frac {\pi }{2}}&n=m\end{cases}}}.$$
The first five ones  are
${{\mathit {U}}_{0}(x)=1}$, ${{\mathit {U}}_{1}(x)=2x}$, ${ {\mathit {U}}_{2}(x)=4x^{2}-1}$,
${{\mathit {U}}_{3}(x)=8x^{3}-4x}$, and ${{\mathit {U}}_{4}(x)=16x^{4}-12x^{2}+1}$. Thus apart from the normalization they coincide with the present polynomials.

\subsection{D-dimensional Fermi-Dirac polynomials}\label{fd}
We seek the set of polynomials orthonormal under the weight defined by the Fermi-Dirac statistical occupation number for $\boldsymbol{u}=0$.
\begin{eqnarray}
\omega(\boldsymbol{\xi}) = \frac{1}{z^{-1}e^{-\frac{\boldsymbol{\xi}^2}{2\theta}}+1}\, \mbox{and} \, \xi_{max}=\infty.
\end{eqnarray}
The parameter $z\equiv e^{\mu/\theta}$ is called the fugacity, where $\mu$  is the chemical potential and $\theta$ is the temperature using the so-called reduced units $(m=k_B=c=\hbar=1)$.
For the Fermi-Dirac weight it holds that,
\begin{eqnarray}
I_{2N}=(2\pi)^{D/2}\theta^{\nu}g_{\nu}\left(z\right), \; \nu \equiv N+D/2, \, \mbox{and} \, g_{\nu} (z) \equiv \int_0^{\infty}dx\, \frac{x^{\nu-1}}{z^{-1}e^x+1}.
\end{eqnarray}
It is interesting to consider special limits where the integral $g_{\nu} (z)$ can be explicitly calculated and the $I_{2N}$  obtained.
One of such limits is when the quantum Fermi-Dirac statistics becomes the classical Maxwell-Boltzmann statistics. This is the small fugacity limit,
\begin{eqnarray}
g_{\nu} (z)= \Gamma(\nu)\Big (z - \frac{z^2}{2^{\nu}}+ \frac{z^3}{3^{\nu}} +\cdots \Big ).
\end{eqnarray}
The other interesting limit is the so-called Sommerfeld limit~\cite{pathria11}, used for the treatment of electrons in metals where the chemical potential (Fermi energy) is much larger than the room temperature. In this limit $\mu/\theta=\ln{z} >>1$, such that the terms of high order in $(\theta/\mu)^{2k}$ can be disregarded.
\begin{eqnarray}\label{sommerfeld}
g_{\nu} (z)= \frac{\Gamma(\nu)}{\Gamma(\nu+1)}(\ln{z})^{\nu}\Big \{ 1+\nu (\nu-1) \frac{\pi^2}{6}\frac{1}{(\ln{z})^2}+\nu (\nu-1)(\nu-2)(\nu-3)\frac{7\pi^4}{360}\frac{1}{(\ln{z})^4}+\cdots \Big \}
\end{eqnarray}

\subsection{D-dimensional Bose-Einstein polynomials}\label{be}
Similarly to the previous case, we seek the set of polynomials orthonormal under a weight which is the Bose-Einstein statistical occupation number.
\begin{eqnarray}
\omega(\boldsymbol{\xi}) = \frac{1}{z^{-1}e^{-\frac{\boldsymbol{\xi}^2}{2\theta}}-1},
\, \mbox{and} \, \xi_{max}=\infty.
\end{eqnarray}
It holds that,
\begin{eqnarray}
I_{2N}=(2\pi)^{D/2}\theta^{\nu}h_{\nu}\left(z\right), \; \alpha\equiv N+D/2, \, \mbox{and} \, h_{\nu} (z) \equiv \int_0^{\infty}dx\, \frac{x^{\nu-1}}{z^{-1}e^x-1}
\end{eqnarray}
The are also special limits here, the first being when the quantum Bose-Einstein  statistics becomes the classical Maxwell-Botlzmann one.
This is the small fugacity limit,
\begin{eqnarray}
h_{\nu} (z)= \Gamma(\nu)\Big (z + \frac{z^2}{2^{\nu}}+ \frac{z^3}{3^{\nu}} +\cdots \Big ).
\end{eqnarray}
Notice that in leading order in $z$, the Fermi-Dirac and the Bose-Einstein polynomials become identical as they reduce to the Maxwell-Boltzmann polynomials at finite temperature. In fact at this limit these polynomials are just scaled versions of the D-dimensional Hermite polynomials.
The other interesting limit is that of negative vanishing fugacity near to the onset of the Bose-Einstein condensate. For this we write $\alpha \equiv - \mu/\theta$, $\alpha \rightarrow 0$. Thus for $z \equiv e^{-\alpha}$, one obtains that,
\begin{eqnarray}
h_{\nu} (z)= \frac{\Gamma(1-\nu)}{\alpha^{1-\nu}}+ \sum_{i=0}^{\infty}\frac{(-1)^i}{i\,\!!}\zeta(\nu-i)\alpha^i,
\end{eqnarray}
where $\zeta(s)$ is the Riemann zeta function, which is defined for ${\mathrm  {Re}}(s)>1$. Thus for $\nu$ integer, the above expression must be replaced by,
\begin{eqnarray}
h_{m} (z)= \frac{(-1)^{m-1}}{(m-1)\,\!!} \big (\sum_{i=1}^{m-1} \frac{1}{i} -\ln{\alpha}\big )\alpha^{m-1}+\sum_{\begin{array}{c} i=0 \\  i \ne m-1 \end{array}}^{\infty}\frac{(-1)^i}{i\, \!!}\zeta(m-i)\alpha^i.
\end{eqnarray}

\subsection{D-dimensional Graphene polynomials}\label{graph}
Graphene is  a two-dimensional sheet of carbon atoms arranged in an hexagonal lattice where electrons move with a relativistic dispersion relation. To deal  with this situation we seek a set of polynomials orthonormal under a weight which the following Fermi-Dirac statistical occupation number for $\boldsymbol{u}=0$\cite{oettinger13}.
\begin{eqnarray}
\omega(\boldsymbol{\xi}) = \frac{1}{z^{-1}e^{\frac{\vert \boldsymbol{\xi}\vert}{\theta}}+1},
\, \mbox{and} \, \xi_{max}=\infty.
\end{eqnarray}
It holds that,
\begin{eqnarray}
I_{2N}= 2^{D+N} \pi ^{\frac{1}{2} (-1+D)} \theta ^{D+2 N} \Gamma\left[\frac{1+D}{2}+N\right] g_{D+2N}(z)\\
\end{eqnarray}
The limit of low doping in graphene corresponds to $\mu \rightarrow 0$ or, equivalently, $z\rightarrow 1$. The doping of graphene can be chemically adjusted for instance~\cite{liu11}.
\begin{eqnarray}\label{graphene}
&&I_{2N}=2^{-N} \left(-2+2^{D+2 n}\right) \pi ^{\frac{1}{2} (-1+D)} \theta ^{D+2 N} \Gamma\left[\frac{1+D}{2}+n\right] \zeta[D+2 N]
\\ \nonumber && +2^{-N}\left(-4+2^{D+2 N}\right) \pi ^{\frac{1}{2} (-1+D)} \theta ^{D+2 N} \Gamma\left[\frac{1+D}{2}+N\right] \zeta[-1+D+2 N] (z-1)
\\ \nonumber &&+2^{-1-N}\pi ^{\frac{1}{2} (-1+D)} \theta ^{D+2 N} \Gamma\left[\frac{1+D}{2}+N\right] \left(\left(-8+2^{D+2 N}\right) \zeta[-2+D+2 N]\right.
\\ \nonumber && \left. -\left(-4+2^{D+2N}\right) \zeta[-1+D+2 N]\right) (z-1)^2+O[z-1]^3
\end{eqnarray}
\subsection{D-dimensional Yukawa polynomials} \label{yuka}
Consider the Yukawa potential of nuclear interactions that contains
the parameter $\mu$ that renders it short ranged.
Assume the Yukawa potential as a weight to obtain orthonormal polynomials.
\begin{eqnarray}
\omega(\boldsymbol{\xi}) = \frac{ e^{-\mu \xi}}{\xi}\, \mbox{and} \, \xi_{max}=\infty.
\end{eqnarray}
To determine the coefficients that enter the polynomials, it is enough to have the integrals,
\begin{eqnarray}
I_{2N}= \frac{\pi^{D/2}}{2^{N-1}} \frac{\Gamma(2N+D-1)}{\Gamma(N+\frac{D}{2})}\frac{1}{\mu^{2N+D-3}}.
\end{eqnarray}
\section{Expansion of a function in general D-dimensional polynomials}\label{expan}
Suppose a  function $f(\boldsymbol{\xi})$ whose argument is the D-dimensional vector $\boldsymbol{\xi}$ such that a small vector $\boldsymbol{u}$ is subtracted from it. We want to expand it in terms of this small correction using the D-dimensional polynomial basis.
\begin{eqnarray}
f(\boldsymbol{\xi} -\boldsymbol{u})= f( \boldsymbol{\xi} ) \sum_{N=0}^{\infty}\frac{1}{N!} \mathcal{A}_{i_1\,i_2 \cdots i_N}(\boldsymbol{u} ) \mathcal{P}_{i_1\,i_2\cdots i_N}( \boldsymbol{\xi}),
\label{feq-complete-exp-eq}
\end{eqnarray}
The coefficients $\mathcal{A}_{i_1\,i_2 \cdots i_N}$ and the polynomials $\mathcal{P}_{i_1\,i_2\cdots i_N}$ only depend on $\boldsymbol{u}$ and $\boldsymbol{\xi}$, respectively.

To help the understanding of the above expansion, we consider the example of a local distribution of charges, $\rho( \boldsymbol{u})$, that interact through a two-body potential, $f(\boldsymbol{u} -\boldsymbol{u^{\prime}})$, such as the Yukawa potential. We want to obtain the potential $v(\boldsymbol{\xi})$ far way from this charge distribution, namely at a distance  ${\boldsymbol{\xi}} \gg {\boldsymbol{u}} $. As the potential produced by each charge at volume $d^D \boldsymbol{u}\, \rho(\boldsymbol{u})$ contributes with potential
$f(\boldsymbol{\xi} -\boldsymbol{u})$ to the total potential, one obtains that,
\begin{eqnarray}
v(\boldsymbol{\xi}) = \int d^D \boldsymbol{u} \,f(\boldsymbol{\xi} -\boldsymbol{u})\rho(\boldsymbol{u}).
\end{eqnarray}
Therefore we propose here a novel expansion in terms of the {\it orthonormal multipoles} $\mathcal{Q}_{i_1\,i_2 \cdots i_N}$, called in this way because of the  orthonormal basis,
\begin{eqnarray}
v(\boldsymbol{\xi})= f( \boldsymbol{\xi} ) \sum_{N=0}^{\infty}\frac{1}{N!} \mathcal{Q}_{i_1\,i_2 \cdots i_N} \mathcal{P}_{i_1\,i_2\cdots i_N}( \boldsymbol{\xi}),\quad \mathcal{Q}_{i_1\,i_2 \cdots i_N}\equiv \int d^D \boldsymbol{u} \, \mathcal{A}_{i_1\,i_2 \cdots i_N}(\boldsymbol{u} )\rho( \boldsymbol{u}).
\label{multipole}
\end{eqnarray}
The key ingredient to obtain these {\it orthonormal multipoles} is that  the potential $f( \boldsymbol{\xi})$ itself enters in the definition of the polynomials $\mathcal{P}_{i_1\,i_2\cdots i_N}( \boldsymbol{\xi})$ as the weight function that renders them orthonormal. To understand this we turn to the general properties of the expansion of Eq.(\ref{feq-complete-exp-eq}). Notice that it is of the form $f({\boldsymbol{\xi}}-{\boldsymbol{u}})  = f({\boldsymbol{\xi}})\cdot S({\boldsymbol{\xi}},{\boldsymbol{u}})$, which 
means that the expansion $S({\boldsymbol{\xi}},{\boldsymbol{u}}\rightarrow 0)\rightarrow 1$. Thus the corrections in ${\boldsymbol{u}}$ are necessarily small.

The key ingredient that renders the above expansion advantageous is the assumption that the weight function that defines the set of orthonormal polynomials is the expanded function itself for $\boldsymbol{u}=0$,
\begin{eqnarray}\label{weight}
\omega(\boldsymbol{\xi})\equiv f( \boldsymbol{\xi} ).
\end{eqnarray}
This means that the function has to fulfill the properties previously required for a weight in order to span a polynomial basis.
In this case the coefficients are readily determined using the orthonormality condition of Eq.(\ref{omeg-feq}).
\begin{eqnarray}
\mathcal{A}_{i_1\,i_2 \cdots i_N}(\boldsymbol{u} )= \int d^D \boldsymbol{\xi} \, f(\boldsymbol{\xi}-\boldsymbol{u}) \mathcal{P}_{i_1\,i_2\cdots i_N}( \boldsymbol{\xi}).\nonumber
\end{eqnarray}
Hence the above expression once fed back to Eq.(\ref{feq-complete-exp-eq}) gives the  completeness relation for the polynomials.
\begin{eqnarray}
f(\boldsymbol{\xi} -\boldsymbol{u})= f( \boldsymbol{\xi} ) \sum_{N=0}^{\infty}\frac{1}{N!} \int d^D \boldsymbol{\xi^{\prime}} \, f(\boldsymbol{\xi^{\prime}}-\boldsymbol{u}) \mathcal{P}_{i_1\,i_2\cdots i_N}( \boldsymbol{\xi^{\prime}}) \mathcal{P}_{i_1\,i_2\cdots i_N}( \boldsymbol{\xi}),
\label{feq-complete-exp-eq2}
\end{eqnarray}
The completness relation is given by
\begin{eqnarray}
 f( \boldsymbol{\xi} ) \sum_{N=0}^{\infty}\frac{1}{N!} \, \mathcal{P}_{i_1\,i_2\cdots i_N}( \boldsymbol{\xi^{\prime}}) \mathcal{P}_{i_1\,i_2\cdots i_N}( \boldsymbol{\xi}) = \delta^{D}(\boldsymbol{\xi^{\prime}}-\boldsymbol{\xi}),
\label{feq-complete-exp-eq3}
\end{eqnarray}
since $f(\boldsymbol{\xi}-\boldsymbol{u})= \int d^D \boldsymbol{\xi^{\prime}} \, \delta^{D}(\boldsymbol{\xi^{\prime}}-\boldsymbol{\xi})f(\boldsymbol{\xi^{\prime}}-\boldsymbol{u})$.\\

In order to determine other properties of the coefficients, we define  $\boldsymbol{\eta}\equiv\boldsymbol{\xi}-\boldsymbol{u}$ to obtain that,
\begin{eqnarray}\label{expansao1}
\mathcal{A}_{i_1\,i_2 \cdots i_N}(\boldsymbol{u} )= \int d^D \boldsymbol{\eta} \, \omega(\boldsymbol{\eta}) \, \mathcal{P}_{i_1\,i_2\cdots i_N}( \boldsymbol{\eta}+\boldsymbol{u}),
\end{eqnarray}
At this point we introduce N=0 polynomial, which is constant, into this expression.
\begin{flalign}
\mathcal{A}_{i_1\,i_2 \cdots i_N}(\boldsymbol{u} )= \frac{1}{c_0}\int d^D \boldsymbol{\eta} \, \omega(\boldsymbol{\eta}) \, \mathcal{P}_0(\boldsymbol{\eta}) \,\mathcal{P}_{i_1\,i_2\cdots i_N}( \boldsymbol{\eta}+\boldsymbol{u}),
\end{flalign}
Hence the determination of the coefficients is reduced to the expansion $\mathcal{P}_{i_1\,i_2\cdots i_N}( \boldsymbol{\eta}+\boldsymbol{u})$ as a sum over polynomials of equal or lower order ($M \le N$), $\mathcal{P}_{i_1\,i_2\cdots i_M}( \boldsymbol{\eta})$.
\begin{eqnarray}
&& \mathcal{P}_{i_1i_2\ldots i_N}(\boldsymbol{\eta}+\boldsymbol{u}) =  \mathcal{U}_0(\boldsymbol{u})\mathcal{P}_{i_1i_2\ldots i_N}(\boldsymbol{\eta}) +\ldots+ \frac{1}{c_0}\mathcal{U}_{i_1i_2\ldots i_N}(\boldsymbol{u})\mathcal{P}_{0}(\boldsymbol{\eta}),
\end{eqnarray}
where $\mathcal{U}_{i_1i_2\ldots i_M}(\boldsymbol{u})$ is a polynomial in $\boldsymbol{u}$ that multiplies the polynomial in $\boldsymbol{\eta}$ of order $N-M$. The sought coefficient is nothing but the term
proportional to $\mathcal{P}_0(\boldsymbol{\eta})$ in this expansion.
The limit $\boldsymbol{u} \rightarrow 0$ in the above expression shows that
$\mathcal{U}_0(0)=1$ while for all other higher order tensors it holds that $\mathcal{U}_{i_1i_2\ldots i_M}(0)=0$. In summary the coefficients become,
\begin{eqnarray}
\mathcal{A}_{i_1i_2\ldots i_N}(\boldsymbol{u})=\frac{1}{c_0^2} \mathcal{U}_{i_1i_2\ldots i_N}(\boldsymbol{u}).
\end{eqnarray}
and $\mathcal{A}_{i_1i_2\ldots i_N}(\boldsymbol{u}=0)= 0$ for $N\geq 1$. \\

Next we directly obtain the coefficients by directly
expanding $\mathcal{P}_{i_1i_2\ldots i_N}(\boldsymbol{\eta}+\boldsymbol{u})$
up to the order N=4.
The expansion of the N=0 polynomial is trivial since $\mathcal{P}_{0}(\boldsymbol{\eta} +\boldsymbol{u})=c_0 $
\begin{eqnarray}
\mathcal{A}_0 = \frac{1}{c_0} \nonumber
\end{eqnarray}
Expanding the N=1 polynomial, $\mathcal{P}_{i_1} (\boldsymbol{\eta} + \boldsymbol{u}) = c_1(\eta_{i_1}+u_{i_1})$, so the coefficient is,
\begin{eqnarray}
\mathcal{A}(\boldsymbol{u}) = \frac{1}{c_0^2}c_1 u_{i_1}.\nonumber
\end{eqnarray}
Expanding the N=2 polynomial, $\mathcal{P}_{i_1i_2}(\boldsymbol{\eta}+\boldsymbol{u}) = c_2 (\eta_{i_1} + u_{i_1})(\eta_{i_2}+u_{i_2}) + [\bar c_2 (\boldsymbol{\eta}+ \boldsymbol{u})^2 + c^{\prime}_2] \delta_{i_1i_2}= \mathcal{P}_{i_1i_2}(\boldsymbol{\eta}) + (c_2/c_1)[u_{i_1}\mathcal{P}_{i_2}(\boldsymbol{\eta}) + u_{i_2}\mathcal{P}_{i_1}(\boldsymbol{\eta})] + (2\bar c_2/c_1) u_{i_3}\mathcal{P}_{i_3}(\boldsymbol{\eta}) \delta_{i_1i_2}+ \mathcal{P}_0(\boldsymbol{\eta})\left[c_2 u_{i_1}u_{i_2} +\bar c_2 \boldsymbol{u}^2 \delta_{i_1i_2} \right]/c_0$, its N=0 coefficient gives that,
\begin{eqnarray}
\mathcal{A}_{i_1i_2}(\boldsymbol{u}) = (c_2 u_{i_1}u_{i_2} + \bar c_2 \boldsymbol{u}^2 \delta_{i_1i_2})I_0, \nonumber
\end{eqnarray}
using that $c_0^2=1/I_0$. Similarly the expansion of the N=3 polynomial, $\mathcal{P}_{i_1i_2i_3}(\boldsymbol{\eta} + \boldsymbol{u})$, contains the N=0 polynomial plus  higher order ones that will be omitted for simplicity.
\begin{eqnarray}
&&\mathcal{P}_{i_1i_2i_3}(\boldsymbol{\eta}+\boldsymbol{u}) = \frac{\mathcal{P}_0}{c_0} \Big \{ \Big[ \frac{c_3 \bar c_2 D c^{\prime}_2}{c_2(c_2 + D\bar c_2)} -  \frac{c_3 c^{\prime}_2}{c_2}- \frac{\bar c_3 D c^{\prime}_2}{(c_2+ D\bar c_2)}+ c^{\prime}_3+\bar c_3\boldsymbol{u}^2 + \frac{2\bar c_3 \bar c_2 D c^{\prime}_2}{c_2(c_2+ D\bar c_2)} - \frac{2\bar c_3 c^{\prime}_2}{c_2} \Big ] \nonumber \\
&& u_{i_4}\delta_{i_1i_2i_3i_4} + c_3u_{i_1}u_{i_2}u_{i_3}  \Big \} + \mathcal{O}(\boldsymbol{\eta})
\end{eqnarray}
Using the equations for the coefficients of polynomials, we have
\begin{eqnarray}
&&\mathcal{P}_{i_1i_2i_3}(\boldsymbol{\eta}+\boldsymbol{u}) = \frac{\mathcal{P}_0}{c_0} \Big\{\Big[\frac{I_2}{I_0}[c_3+\bar c_3(D+2)] + \bar c_3\boldsymbol{u}^2 + c^{\prime}_3\Big ]u_{i_4}\delta_{i_1i_2i_3i_4} +  c_3u_{i_1}u_{i_2}u_{i_3}  \Big \} + \mathcal{O}(\boldsymbol{\eta}).
\end{eqnarray}
Therefore the third order coefficient of the expansion is
\begin{eqnarray}
&& \mathcal{A}_{i_1i_2i_3}(\boldsymbol{u}) = I_0\Big \{\Big [\frac{I_2}{I_0}[c_3+\bar c_3(D+2)]  + \bar c_3\boldsymbol{u}^2 + c^{\prime}_3\Big]u_{i_4}\delta_{i_1i_2i_3i_4} +  c_3u_{i_1}u_{i_2}u_{i_3}  \Big \}, \nonumber
\end{eqnarray}
Expanding the N=4 polynomial $\mathcal{P}_{i_1i_2i_3i_4}(\boldsymbol{\eta}+\boldsymbol{u})$ in terms of  $\boldsymbol{\eta}$ is a laborious task as terms in $\boldsymbol{\eta}$ must be expressed again as functions of polynomials of order N. Nevertheless we only seek the N=0 term and some considerations can be applied to simplify the task. For instance, the odd terms ($\eta_{i_1}$, $\eta_{i_1}\boldsymbol{\eta}^2$, $\eta_{i_1}\eta_{i_2}\eta_{i_3}$) do not contribute to the calculation of $\mathcal{A}_{i_1i_2i_3i_4}$ and one can take that $\boldsymbol{\eta}^2 = DI_2/I_0 + \mathcal{O}(\boldsymbol{\eta})$. After some algebra, we have:
\begin{eqnarray}
&&\mathcal{P}_{i_1i_2i_3i_4}(\boldsymbol{\eta}+\boldsymbol{u}) = \frac{\mathcal{P}_0}{c_0} \Big \{c_4 u_{i_1}u_{i_2}u_{i_3}u_{i_4} + \Big[ \frac{I_2}{I_0}[c_4+\bar c_4(D+4)]+ c_4' + \bar c_4 \boldsymbol{u}^2\Big] (\delta_{i_1i_2}u_{i_3}u_{i_4} + \delta_{i_1i_3}u_{i_2}u_{i_4}\nonumber \\
&&+ \delta_{i_1i_4}u_{i_2}u_{i_3}+ \delta_{i_2i_3}u_{i_1}u_{i_4}+\delta_{i_2i_4}u_{i_1}u_{i_3}+\delta_{i_3i_4}u_{i_1}u_{i_2}) + \Big [ 2\bar c_4 \frac{I_2}{I_0}\boldsymbol{u}^2 + d^{\prime}_4 \boldsymbol{u}^2  + 2D\frac{I_2}{I_0} \bar d_4 \boldsymbol{u}^2 + 4\frac{I_2}{I_0}\bar d_4 \boldsymbol{u}^2 \nonumber \\
&&+ \boldsymbol{u}^4 \bar d_4\Big ]\delta_{i_1i_2i_3i_4} \Big \} + \mathcal{O}(\boldsymbol{\eta}).
\end{eqnarray}
Finally, the N=4 coefficient is
\begin{eqnarray}
&&\mathcal{A}_{i_1i_2i_3i_4}(\boldsymbol{u}) = I_0 \Big\{c_4 u_{i_1}u_{i_2}u_{i_3}u_{i_4} + \Big[\frac{I_2}{I_0}[c_4+\bar c_4(D+4)]+ c_4' + \bar c_4 \boldsymbol{u}^2\Big ](\delta_{i_1i_2}u_{i_3}u_{i_4} + \delta_{i_1i_3}u_{i_2}u_{i_4}+ \delta_{i_1i_4}u_{i_2}u_{i_3}+ \nonumber \\
&&\delta_{i_2i_3}u_{i_1}u_{i_4}+\delta_{i_2i_4}u_{i_1}u_{i_3} +\delta_{i_3i_4}u_{i_1}u_{i_2}) + \Big[ 2\bar c_4 \frac{I_2}{I_0}\boldsymbol{u}^2 +  d^{\prime}_4 \boldsymbol{u}^2 +2D\frac{I_2}{I_0} \bar d_4 \boldsymbol{u}^2 + 4\frac{I_2}{I_0}\bar d_4 \boldsymbol{u}^2 + \boldsymbol{u}^4 \bar d_4\Big ] \delta_{i_1i_2i_3i_4}\Big\}
\end{eqnarray}
We summarize the coefficients below, obtained after some additional algebraic manipulation. Notice that they are functions of the integrals $I_{2N}$ previously defined.
\begin{eqnarray}
&&\mathcal{A}_0(\boldsymbol{u}) = I_0 c_0, \label{A0}\\
&&\mathcal{A}_{i_1}(\boldsymbol{u})=I_0 c_1\,u_{i_1},\label{A1}\\
&&\mathcal{A}_{i_1 i_2}(\boldsymbol{u})= I_0 \big ( c_2 u_{i_1} u_{i_2} + {\bar c}_2 \boldsymbol{u}^2 \,\delta_{i_1 i_2} \big ),\label{A2}\\
&&\mathcal{A}_{i_1 i_2 i_3}(\boldsymbol{u})=I_0\big \{ c_3 \,u_{i_1} u_{i_2}u_{i_3}  +\big [ {c^\prime}_3 \big(1-J_2 \big)+{\bar c}_3\boldsymbol{u}^2\big]\big (u_{i_1} \delta_{i_2 i_3} + u_{i_2}\delta_{i_1 i_3} +u_{i_3} \delta_{i_1 i_2}\big ) \big \}, \quad \mbox{and},  \label{A3}\\
&&\mathcal{A}_{i_1 i_2 i_3 i_4}(\boldsymbol{u})=I_0\Big\{{c}_4 \,u_{i_1} u_{i_2}u_{i_3} u_{i_4}+ \big [\big( 1-J_2J_4\big){c^\prime}_4+{\bar c}_4 \boldsymbol{u}^2\big]\big ) \big (u_{i_1}u_{i_2}\delta_{i_3 i_4} + u_{i_1}u_{i_3}
\delta_{i_2 i_4}+u_{i_1}u_{i_4}\delta_{i_2 i_3}+ \nonumber \\
&& u_{i_2}u_{i_3} \delta_{i_1 i_4}+ u_{i_2}u_{i_4} \delta_{i_2 i_3}+u_{i_3}u_{i_4}\delta_{i_1 i_2}\big )+
\big[ \big(2\frac{I_2}{I_0}\big( \bar c_4+(D+2)\bar d_4\big) +d^\prime_4\big ) \boldsymbol{u}^2+\bar d_4 \boldsymbol{u}^4  \big ]\,\delta_{i_1 i_2 i_3 i_4}\Big \}.\label{A4}
\end{eqnarray}
The contraction between tensors
$\mathcal{A}_{i_1i_2\ldots i_N}$ and $\mathcal{P}_{i_1i_2\ldots i_N}$
up to N=4 order are obtained below.
\begin{eqnarray}
\mathcal{A}_0 \mathcal{P}_0 = 1
\end{eqnarray}
\begin{eqnarray}
\mathcal{A}_{i_1} \mathcal{P}_{i_1} = \frac{I_0}{I_2} (\boldsymbol{\xi}\cdot\boldsymbol{u})
\end{eqnarray}
\begin{eqnarray}
\mathcal{A}_{i_1i_2} \mathcal{P}_{i_1i_2} = \frac{I_0}{I_4}(\boldsymbol{\xi}\cdot\boldsymbol{u})^2 - \frac{I_0(\Delta_2^2-1)}{I_4D}\boldsymbol{u}^2\boldsymbol{\xi}^2 - \frac{I_2}{I_4}\Delta_2^2\boldsymbol{u}^2
\end{eqnarray}
\begin{eqnarray}
&&\mathcal{A}_{i_1i_2i_3} \mathcal{P}_{i_1i_2i_3} = I_0 (\boldsymbol{\xi}\cdot\boldsymbol{u})\Big [3(1-J_2)\frac{J_4}{I_2}(D+2)\Delta_4^2 - 3\frac{J_4}{I_4}\Delta_4^2 \boldsymbol{u}^2- 3 (1-J_2)\frac{J_4}{I_4}\Delta_4^2 \boldsymbol{\xi}^2  + 3\frac{\Delta_4^2-1}{I_6(D+2)}\boldsymbol{\xi}^2\boldsymbol{u}^2 +  \frac{1}{I_6}(\boldsymbol{\xi}\cdot\boldsymbol{u})^2\Big] \nonumber \\
\end{eqnarray}
\begin{eqnarray}
&&\mathcal{A}_{i_1i_2i_3i_4} \mathcal{P}_{i_1i_2i_3i_4} = I_0 \Big\{ c_4^2 (\boldsymbol{\xi}\boldsymbol{u})^2 + 6c_4(c_4' + \bar c_4 \xi^2) u^2(\boldsymbol{\xi} \boldsymbol{u})^2+ 3c_4 u^4 (d_4 + d_4'\xi^2 + \bar d_4 \xi^4) + 6\big[ \frac{I_2}{I_0}(c_4+ \nonumber \\
&&\bar c_4(D+4))+ c_4' + \bar c_4 u^2 \big]\big[ c_4 \xi^2 (\boldsymbol{\xi}\cdot \boldsymbol{u})^2 + (c^{\prime}_4 +\bar c_4 \xi^2) [\xi^2 u^2 + (\boldsymbol{\xi} \boldsymbol{u})^2(D+4)]  + (d_4 +d_4'\xi^2 + \bar d_4 \xi^4)u^2(D+2)\big] + \nonumber \\
&&3\big[ u^2 \frac{I_2}{I_0} 2(\bar c_4 + D \bar d_4 + 2\bar d_4)+ d_4' u^2 + \bar d_4 u^4 \big] [c_4\xi^4 +  2(c_4' + \bar c_4 \xi^2) \xi^2 (D+2) + (d_4 + d_4' \xi^2 + \bar d_4 \xi^4)D (D+2)] \Big\} \nonumber \\
\end{eqnarray}

\section{Conclusions}\label{concl}
We propose here D-dimensional symmetric tensor polynomials orthonormal under a general weight. We show that the number of coefficients of the first five ones (N=0 to 4) matches exactly the number of equations that stem from their orthonormalization, which allows for their obtainment as  functions of the integrals $I_N$'s. In Statistical Mechanics it is well-known that the D-dimensional Hermite polynomials are a key element to solve the Boltzmann equation for classical particles, which satisfy the Maxwell-Boltzmann statistics~\cite{kremer10, philippi06, coelho16-2}. The present generalized polynomials are applicable to semi-classical fluids where the particles obey the Bose-Einstein and Fermi-Dirac statistics~\cite{coelho14,coelho16}.
The proposed generalized polynomials allows for the definition of orthonormal multipoles as shown in case of the Yukawa potential. We foresee many other applications because the proposed polynomials take into account the expanded function as the weight that render them orthonormal in the D-dimensional space.

\begin{acknowledgments}
R. C. V. Coelho thanks to FAPERJ and to the European Research Council (ERC) Advanced Grant 319968-FlowCCS for the financial support and to Hans J. Herrmann for the kind hospitality at ETH Z\"{u}rich.
\end{acknowledgments}

\appendix
\section{Tensorial Identities involving the $I_N$} \label{appendixb}
Here we obtain especial formulas derived from the definition of the $I_N$'s
given in Eq.(\ref{i2n-cont}) and obtained from tensorial contractions of the tensor $\xi_{i_1}\cdots \xi_{i_{N}}$, and, consequently, from the $\delta_{i_1\cdots i_{N}}$.
\begin{eqnarray}
&& \int d^D \boldsymbol{\xi} \, \omega( \boldsymbol{\xi} ) = I_{0}. \nonumber \end{eqnarray}
\begin{eqnarray}
&& \int d^D \boldsymbol{\xi} \, \omega( \boldsymbol{\xi} ) \, \xi_{i_1} \xi_{j_1}=  I_{2}\,\delta_{i_1 j_1}, \nonumber \\
&& \int d^D \boldsymbol{\xi} \, \omega( \boldsymbol{\xi} ) \, \xi^2 =  D \, I_{2}.
\end{eqnarray}
\begin{eqnarray}
&& \int d^D \boldsymbol{\xi} \, \omega( \boldsymbol{\xi} ) \, \xi_{i_1} \xi_{i_2} \xi_{j_1}\xi_{j_2} =  I_{4}\,\delta_{i_1 j_1 i_2 j_2}, \nonumber \\
&& \int d^D \boldsymbol{\xi} \, \omega( \boldsymbol{\xi} ) \, \xi_{i_1} \xi_{i_2} \xi^2 = (D+2) I_{4}\,\delta_{i_1 j_1}, \\
&& \int d^D \boldsymbol{\xi} \, \omega( \boldsymbol{\xi} ) \,  \xi^4 = (D+2)D I_{4}.
\end{eqnarray}
\begin{eqnarray}
&& \int d^D \boldsymbol{\xi} \, \omega( \boldsymbol{\xi} ) \, \xi_{i_1} \xi_{i_2}\xi_{i_3} \xi_{j_1}\xi_{j_2}\xi_{j_3} =  I_{6}\,\delta_{i_1 j_1 i_2 j_2 i_3 j_3}, \nonumber \\
&& \int d^D \boldsymbol{\xi} \, \omega( \boldsymbol{\xi} ) \, \xi_{i_1} \xi_{i_2} \xi_{j_1}\xi_{j_2}\xi^2 =  (D+4) I_{6} \,\delta_{i_1 j_1 i_2 j_2},\\
&& \int d^D \boldsymbol{\xi} \, \omega( \boldsymbol{\xi} ) \, \xi_{i_1} \xi_{i_2} \xi^4 = (D+4)(D+2) I_{6} \,\delta_{i_1 j_1},\\
&& \int d^D \boldsymbol{\xi} \, \omega( \boldsymbol{\xi} ) \, \xi^4 = (D+4)(D+2)D I_{6} .
\end{eqnarray}
\begin{eqnarray}
&& \int d^D \boldsymbol{\xi} \, \omega( \boldsymbol{\xi} ) \, \xi_{i_1} \xi_{i_2}\xi_{i_3}\xi_{i_4} \xi_{j_1}\xi_{j_2}\xi_{j_3}\xi_{j_4} =  I_{8}\,\delta_{i_1 j_1 i_2 j_2 i_3 j_3 i_4 j_4}, \nonumber \\
&& \int d^D \boldsymbol{\xi} \, \omega( \boldsymbol{\xi} ) \, \xi_{i_1} \xi_{i_2}\xi_{i_3} \xi_{j_1}\xi_{j_2}\xi_{j_3}\xi^2 =  (D+6)I_{8}\,\delta_{i_1 j_1 i_2 j_2 i_3 j_3}, \nonumber  \\
&& \\
&& \int d^D \boldsymbol{\xi} \, \omega( \boldsymbol{\xi} ) \, \xi_{i_1} \xi_{i_2} \xi_{j_1}\xi_{j_2}\xi^4 =  (D+6)(D+4)I_{8}\,\delta_{i_1 j_1 i_2 j_2}, \nonumber \\
&& \\
&& \int d^D \boldsymbol{\xi} \, \omega( \boldsymbol{\xi} ) \, \xi_{i_1} \xi_{j_1}\xi^6 =  (D+6)(D+4)(D+2)I_{8}\,\delta_{i_1 j_1}, \nonumber \\
&& \\
&& \int d^D \boldsymbol{\xi} \, \omega( \boldsymbol{\xi} ) \, \xi^8 =  (D+6)(D+4)(D+2)DI_{8}.
\end{eqnarray}

\bibliography{reference}

\end{document}